\documentclass[11pt]{article}
\usepackage{mymacros}

\title{ Probing the Chaos to Integrability
Transition in Double-Scaled SYK }
\author[a,1]{Sergio E. Aguilar-Gutierrez\,\orcidlink{0000-0003-0308-0061},\note{Corresponding authors.}}
\author[b, c, d]{Rathindra Nath Das\,\orcidlink{0000-0002-4766-7705},}
\author[b]{Johanna Erdmenger\,\orcidlink{0000-0003-4776-4326},}
\author[e,1]{and Zhuo-Yu Xian\,\orcidlink{0000-0002-2245-0059}}
\affiliation[a]{Qubits and Spacetime Unit, Okinawa Institute of Science and Technology Graduate University\footnote{\begin{CJK}{UTF8}{min}沖縄科学技術大学院大学\end{CJK}}, Onna, Okinawa 904 0495, Japan}
\affiliation[b]{Institute for Theoretical Physics and Astrophysics and Würzburg-Dresden Cluster of Excellence ctd.qmat, Julius-Maximilians-Universität Würzburg, Am Hubland, 97074 Würzburg, Germany}

\affiliation[c]{Department of Particle Physics and Astrophysics, Weizmann Institute of Science, Rehovot 7610001, Israel}
\affiliation[d]{MIT Center for Theoretical Physics—a Leinweber Institute, Massachusetts Institute of Technology, 77 Massachusetts Ave., Cambridge, MA 02139}

\affiliation[e]{Department of Physics, Freie Universit\"at Berlin, Arnimallee 14, DE-14195 Berlin, Germany}
\emailAdd{sergio.ernesto.aguilar@gmail.com}
\emailAdd{das.rathindranath@uni-wuerzburg.de}
\emailAdd{zhuo-yu.xian@fu-berlin.de}

\abstract{We investigate how a thermodynamical first-order phase transition affects the dynamical chaotic behaviour of a given model.
To this effect, we  analyze the model of Berkooz, Brukner, Jia and Mamroud that interpolates between the double-scaled SYK model and an integrable chord Hamiltonian. This model exhibits a first-order transition, characterized by a kink in the free energy, between the chaotic and quasi-integrable phases, with the branch of subdominant saddles interpolating between them.
We characterize the dynamical behavior across the phase diagram using the chord number, Krylov complexity, and operator size. 
The chord number, which is proportional to the Krylov state complexity in the classical limit, exhibits a discontinuous transition from linear to quadratic growth at the transition point.
Similarly, the Krylov operator complexity and the operator size, as scrambling diagnostics, exhibit discontinuous transitions from exponential to quadratic growth. We also discuss a possible holographic interpretation of the model.}

\begin{document}

\maketitle

\newpage
\section{Introduction}
Phase transitions play a central role in understanding how many-body systems reorganize under changes in thermodynamic control parameters, and it is natural to ask how signatures of quantum chaos behave across such transitions. There has been considerable interest in exploring various measures of quantum chaos including early-time measures such as out-of-time-ordered correlators (OTOCs)\cite{aleiner1996divergence,Rozenbaum:2016mmv} and Krylov operator complexities \cite{Parker:2018yvk}; and late-time measures such as the spectral form factor \cite{Cotler:2016fpe}, level spacing spectral statistics \cite{Dyson:1962es,Dyson:1962oir,wigner1993characteristic}, and Krylov state complexities \cite{Balasubramanian:2022tpr,Erdmenger:2023wjg}; see \cite{Rabinovici:2025otw,Baiguera:2025dkc,Nandy:2024evd} for reviews. In particular, several works have found that Krylov complexity is a useful measure to characterize systems transitioning between chaotic and integrable properties \cite{He:2022ryk,Balasubramanian:2024ghv,Camargo:2024deu,Rabinovici:2022beu,Alishahiha:2024vbf,Rabinovici:2021qqt,Baggioli:2024wbz,Scialchi:2023bmw,Bhattacharjee:2024yxj,Camargo:2023eev,Baggioli:2025knt}.

In this work we consider the model of Berkooz, Brukner, Jia and Mamroud (BBJM) \cite{Berkooz:2024ofm,Berkooz:2024evs} as a concrete framework to study a phase transition between integrable and chaotic behaviour (see also \cite{Berkooz:2024ifu,Almheiri:2024xtw}). In its simplest version, the BBJM model is an interpolation between two Hamiltonians in a double-scaling limit\footnote{This can be extended by including chords with ``flavours'' \cite{Gao:2024lem,Berkooz:2024ifu}.}, in which each theory can be described by the techniques of chord diagrams, explained below, namely
\begin{equation}\label{eq:chaos-integrability interpolating Hamiltonian}
\hat{H}=\nu \hat{H}_{1}+\kappa \hat{H}_{2}\,,\quad \abs{\nu}^2+\abs{\kappa}^2=1\,.
\end{equation}

\begin{figure}
    \centering
    {\includegraphics[height=0.42\textwidth]{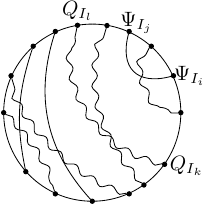}}
    \caption{Example of a chord diagram
    representing the computation of Hamiltonian moments \eqref{eq:trace H}, where the trace is represented by a circle. There are two types of chords, both corresponding to the contraction between either two strings of fermions (solid curves connecting vertices, such as $\Psi_{I_i}$ and $\Psi_{I_j}$); or two strings of bosonic operators (wavely curves connecting vertices, such as $Q_{I_k}$, $Q_{I_l}$).}
    \label{fig: chord diagram}
\end{figure}

We take $\hat{H}_1$ to be the chord Hamiltonian of the double-scaled SYK (DSSYK) model \cite{Berkooz:2018qkz,Berkooz:2018jqr}, reviewed in \cite{Berkooz:2024lgq}\footnote{Quantum groups and von Neumann algebras in this and other closely related systems have been explored in \cite{Berkooz:2018jqr,Berkooz:2022mfk,Blommaert:2023opb, Almheiri:2024ayc,Belaey:2025ijg, vanderHeijden:2025zkr, Schouten:2025tvn,Xu:2024hoc,Cao:2025pir}.} while $\hH_2$ is an integrable Hamiltonian that we specify below. 
A chord refers to a Wick contraction between pairs of Majorana fermion strings $\Psi_I$, in the calculation of Hamiltonian moments, such as
\begin{equation}\label{eq:trace H}
\expval{\tr H_{\rm SYK}^k}=\rmi^{kp/2}\sum_{I_1\dots I_k}\expval{J_{I_1}\cdots J_{I_k}}\tr{\Psi_{I_1}\cdots\Psi_{I_k}}~,
\end{equation}
where the pairwise contraction follows from the Gaussian distribution for the random coefficients $J_I$. There are similar Wick contractions occurring in the integrable model, which has its own set of random couplings. We present an example of the chord diagram including both types of chords in the Fig.~\ref{fig: chord diagram}.
An important insight from \cite{Berkooz:2018jqr,Berkooz:2018qkz} is that the calculation of the averaged Hamiltonian moments in the microscopic theory \eqref{eq:trace H} can be performed in terms of an auxiliary quantum system, which is described by the corresponding chord Hamiltonian, namely
\begin{equation}
\bra{0}\hH^k\ket{0}=\expval{\tr H_{\rm SYK}^k}~,
\end{equation}
where $\ket{0}$ corresponds to the tracial state in the auxiliary system.
Note that in the DSSYK model, 
maximal chaos is well-established 
from different chaos measures \cite{Aguilar-Gutierrez:2025pqp,Xu:2024gfm,Aguilar-Gutierrez:2025mxf,Lin:2023trc}.

Meanwhile $\hat{H}_2$ is an integrable chord Hamiltonian \cite{Berkooz:2024ofm}, for instance a commuting SYK model \cite{Gao:2023gta} in the double-scaling regime \cite{Gao:2024lem,Berkooz:2024ofm,Berkooz:2024evs,Almheiri:2024xtw}, or an integrable spin chain\footnote{A more general formalism with an arbitrary number of chord Hamiltonians was later developed in \cite{Berkooz:2024ifu,Gao:2024lem}, which we review in App.\,\ref{sapp:multifield}.}, that admits a similar chord diagrammatic expansion as the one presented above.  
We allow for crossings between integrable chords and chaotic chords (see Fig.\,\ref{fig: chord diagram}).

In addition, recent deve\-lopments indicate that the holographic dual of the DSSYK model and the commuting DSSYK are sine dilaton gravity \cite{Blommaert:2024ydx,Blommaert:2023opb,Blommaert:2024whf,Blommaert:2024ydx,Blommaert:2025avl,Blommaert:2025rgw,Bossi:2024ffa,Blommaert:2025eps,Cui:2025sgy} and its flat space limit \cite{Blommaert:2024whf},  respectively\footnote{Other proposals for the holographic dual to the DSSYK model include three-dimensional de Sitter (dS$_3$) space \cite{Narovlansky:2023lfz,Verlinde:2024znh,Verlinde:2024zrh,Narovlansky:2025tpb,Blommaert:2025eps,HVtalks,Aguilar-Gutierrez:2024nau}, and dS$_2$ Jackiw-Teitelboim (JT) gravity \cite{JACKIW1985343,TEITELBOIM198341} as a dimensional reduction of dS$_3$ space \cite{Susskind:2021esx,Susskind:2022bia,Susskind:2023hnj,Lin:2022nss,Rahman:2022jsf,Rahman:2023pgt,Rahman:2024iiu,Rahman:2024vyg,Sekino:2025bsc,Miyashita:2025rpt,Tietto:2025oxn}, among others \cite{Milekhin:2023bjv,Okuyama:2025hsd,Yuan:2024utc,Aguilar-Gutierrez:2024nau,Gaiotto:2024kze}. These proposals may be related among each other, and including sine dilaton gravity \cite{Blommaert:2025eps,Aguilar-Gutierrez:2025hty,Aguilar-Gutierrez:2024oea,Aguilar-Gutierrez:2025otq,Heller:2025ddj}.}. Krylov complexity is a central entry in the holographic dictionary of the DSSYK model. It   has a geometric realization as a wormhole length in the dual bulk theory \cite{Rabinovici:2023yex,Heller:2024ldz,Lin:2023trc,Aguilar-Gutierrez:2025hty,Aguilar-Gutierrez:2025mxf,Aguilar-Gutierrez:2025pqp,Xu:2024hoc,Balasubramanian:2024lqk,Heller:2025ddj,Aguilar-Gutierrez:2025sqh,Fu:2025kkh}.  It is important to develop this program further, in particular in view of understanding the consequences of the integrable deformations in the DSSYK  for the holographic dictionary beyond standard the anti-de Sitter (AdS)/conformal field theory (CFT) correspondence.

In addition to the remarkable progress in understanding the properties of the BBJM model \eqref{eq:chaos-integrability interpolating Hamiltonian}, here we consider dynamical measures of chaos across the phase diagram. 
Following the terminology of \cite{Berkooz:2024evs,Berkooz:2024ofm}, we refer to the phase continuously connected to the purely chaotic system at $\kappa = 0$ as the \emph{chaotic phase}, and to the phase continuously connected to the purely integrable system at $\nu = 0$ as the \emph{quasi-integrable phase}. At this stage, these names are adopted as labels reflecting their limiting behavior, rather than as a priori dynamical diagnoses of chaos/integrability.

While there is a clear distinction between the two phases on either side of the transition point, as captured by the free energy exhibiting a kink, it remains unclear whether this thermodynamic distinction is reflected in dynamical measures of quantum chaos \cite{Berkooz:2024evs,Berkooz:2024ofm}. We therefore ask:
\begin{quote}
\emph{Do the quasi-integrable and chaotic phases exhibit distinct signatures of quantum chaos, separated by the first-order phase transition?}
\end{quote}
Our result will answer affirmatively this question, there are different scrambling measures that detect the phase transition between quasi-integrable and chaotic phases. To characterize quantum chaos in the BBJM model in the double-scaling limit, we focus on scrambling rather than spectral statistics. 
On the one hand, scrambling describes how local information spreads throughout the system. Systems that exhibit fast scrambling show exponential growth of OTOCs or Krylov operator complexity before the scrambling time. These quantities can be computed in the double-scaling limit. Although scrambling does not necessarily imply quantum chaos, where exponential growths of the OTOCs and Krylov complexity can arise from unstable saddle points or local maxima even in integrable or non-chaotic systems \cite{Xu:2019lhc,Hashimoto:2020xfr,Bhattacharjee:2022vlt,Trunin:2023xmw}, it remains important to examine scrambling in the BBJM model, in order to test whether the phase classification is reflected in distinct dynamical behavior\footnote{There are certain examples in the literature when Krylov operator complexity in quantum mechanical systems is not a reliable measure of chaos \cite{Chapman:2024pdw,Bhattacharjee:2022vlt,Huh:2023jxt} at early-times,in contrast to other definitions of quantum chaos, e.g.~in terms of level spacing statistics. Nevertheless, Krylov operator complexity can still be reliable at sufficiently late-time regimes \cite{Aguilar-Gutierrez:2025hbf}.
However, we will only probe time scales comparable or below the scrambling time in this system.}.

On the other hand, level-spacing statistics characterize quantum chaos through random-matrix behavior of energy levels, as reflected in the spectral form factor. 
Such statistics are well-defined for systems with discrete energy spectra, such as a single realization of the SYK model or the commuting SYK model at finite $N$, i.e., a finite number of Majorana fermions. 
However, we work with an auxiliary system describing ensemble-averaged observables of the SYK \cite{kitaevTalks,Sachdev:1992fk} and commuting SYK \cite{Gao:2023gta} models in the double-scaling limit with $N \to \infty$. In this limit, the energy spectrum becomes continuous, and spectral measures of chaos are not well-defined\footnote{In the context of AdS/CFT, early-time chaotic observables (such as OTOCs and Krylov complexity) capture the $N \to \infty$ limit of the boundary theory, corresponding to disk topology in the bulk Euclidean path integral. In contrast, late-time features such as the ramp and plateau in the spectral form factor, as well as the saturation of Krylov complexity, require perturbative and non-perturbative finite-$N$ effects, corresponding to wormholes and higher-genus contributions in the bulk.}. 

For this reason, we test the phase classification through scrambling diagnostics, such as Krylov operator complexity, operator size, and OTOCs, and pose the following questions:
\begin{itemize}
    \item Does the quasi-integrable phase exhibit the absence of fast scrambling, e.g., polynomial growth of Krylov complexity and operator size? 
    \item Does the chaotic phase exhibit fast scrambling, characterized by exponential growth of these quantities?
    \item Do scrambling diagnostics exhibit a sharp change at the first-order transition point?
\end{itemize}

In addition, the behavior of scrambling across the first-order transition in the thermodynamic limit warrants further investigation from a many-body perspective. 
A first-order phase transition does not necessarily imply a qualitative change in chaotic properties. For example, charged black holes can undergo a first-order phase transition between small and large black hole phases \cite{Chamblin:1999tk,Chamblin:1999hg}, yet both phases saturate the Lyapunov bound \cite{Awal:2025irl,Shukla:2024tkw}. 
Moreover, scrambling diagnostics are crucial for assessing whether the model can be realized in table-top experiments (see, e.g., \cite{Jafferis:2022crx}), as envisioned in \cite{Berkooz:2024evs,Berkooz:2024ofm}.

In this work, we address this question by examining the integrable-chaotic chord Hamiltonian model proposed in \cite{Berkooz:2024evs,Berkooz:2024ofm}, using different measures of scrambling. 
We focus on (i) deriving the Krylov basis for spread complexity using the thermo-field double (TFD) as the reference state in the Lanczos algorithm, (ii) Krylov operator complexity, and (iii) operator size at finite temperature for chaotic matter chord operators. 
These diagnostics allow us to test whether the quasi-integrable and chaotic phases exhibit the expected scrambling behavior of integrable and chaotic systems, respectively.

We consider the classical limit of the BBJM model and numerically solve the corresponding Liouville equations for imaginary time. By unfolding the free-energy kink, we track the solutions across the chaotic, subdominant, and quasi-integrable branches of saddles. The solutions change continuously when changing $\kappa$, even at low temperatures. By solving the Liouville equations for real time, we obtain the time evolution of chord number, Krylov operator complexity and operator size. Comparing the early-time behavior of Krylov operator complexity and operator size across the different branches of saddles, we find exponential growth for the chaotic saddles, quadratic growth for the quasi-integrable saddles, and intermediate behavior for the subdominant saddles. However, across the first-order phase transition between the chaotic and quasi-integrable phases, the system bypasses the subdominant saddles and the growth behavior of both Krylov operator complexity and operator size instead exhibits a discontinuous jump at the transition point. Thus, the transition between quasi-integrable and chaotic phases is indeed visible by a distinct jump in the scrambling measures.

\paragraph{Outline}
In Sec.\,\ref{eq:background}, we provide a brief introduction to the integrable-chaotic chord Hamiltonian model in \cite{Berkooz:2024evs,Berkooz:2024ofm} and investigate the growth of chord numbers across the integrable-chaotic transition.  
In Sec.\,\ref{sec:krylov basis spread}, we explore the Krylov basis with respect to the infinite-tempera\-ture TFD as the reference state. 
In Sec.\,\ref{sec:Krylov operators}, we derive two-point correlation functions and the corresponding Krylov operator complexity for chaotic operators. 
In Sec.\,\ref{sec:OTOCs}, we analyze the operator size growth for the same type of operators at finite temperature. 
In Sec.\,\ref{sec:discussion}, we conclude summarizing and interpreting the results; besides mentioning some directions for future research. 

We also provide additional details in App.\,\ref{app:background} to complement the background material in Sec.\,\ref{eq:background}; App.\,\ref{app:alt exp} and App.\,\ref{ssec:deviations krylov basis} technical details regarding the Krylov basis in Sec.\,\ref{ssec:special case}.

\section{The integrable-chaotic model and its phase transition}\label{eq:background}

In Sec.\,\ref{ssec:chord Hilbert space}, we review the BBJM\footnote{Additional supplementary material regarding Krylov complexity for states and operators, OTOCs, and about ensemble-averaged doubled-scaled models can be found in App.\,\ref{app:background}.} model and its chord Hilbert space. In Sec.\,\ref{ssec:correlation}, we review its effective action and phase transition in the classical limit. In Sec.\,\ref{ssec:evol phases}, we present new results regarding the evolution of the chord numbers.

\subsection{The BBJM model and its chord Hilbert space}\label{ssec:chord Hilbert space}
The construction of the chord Hilbert space of the integrable-chaotic chord Hamiltonian \eqref{eq:chaos-integrability interpolating Hamiltonian} follows in a similar way to \cite{Berkooz:2018jqr}, which is reviewed in App.\,\ref{sapp:DSSYK basic}. Let us briefly describe the DSSYK model in some detail. The SYK model is a strongly interacting system of $N$ Majorana fermions ${\psi}_{i}$ obeying $\qty{{\psi}_{i},~{\psi}_{j}}=2\delta_{ij}$ in $(0+1)$-dimensions with all to all $p$ body interactions (see \cite{Sachdev:1992fk,kitaevTalks,Cotler:2016fpe,Maldacena:2016upp,Saad:2018bqo,Maldacena:2018lmt,Jensen:2016pah,Polchinski:2016xgd,Chowdhury:2021qpy} among many others) 
\begin{equation}\label{eq:SYK Hamiltonian}
    {H}^{(p)}_{\rm SYK}=\rmi^{p/2}\sum_{I}J_I{\Psi}_I\,,
\end{equation}
where $I$ is a collective index indicating $1\leq i_1\leq i_2\leq\dots\leq i_p\leq N$, and we represent ${\Psi}_I:={\psi}_{i_1}\dots{\psi}_{i_p}$; $J_I:=J_{i_1,\dots,~i_p}$ are random coupling constants which obey the following Gaussian distribution
\begin{equation}\label{eq:GEA J}
\expval{J_I}=0\,,\quad
\expval{J_IJ_{I'}}=\delta_{II'}{\mathcal{J}^2}
\binom{N}{p}^{-1}\,,
\end{equation}
where $\mathcal{J}$ is a constant independent of $N$ and $p$. The double scaling refers to
\begin{equation}\label{eq:double scaling}
    N,~ p \rightarrow \infty\,,\quad \lambda \equiv \frac{2p^2}{N}~ \text{fixed}\,.
\end{equation}
Using the double scaling limit, it can be shown that the ensemble-averaged expectation values of powers of Hamiltonian (\ref{eq:SYK Hamiltonian}) are exactly reproduced by constructing an auxiliary Hilbert space $\mathcal{H}={\rm span}\qty{\ket{n}}$, where $\ket{n}$ represent chord-number states (see \cite{Berkooz:2024evs} for a review), and the Hamiltonian acting on $\mathcal{H}$ is denoted $\hH_1$.

The commuting SYK model was first proposed in \cite{Gao:2023gta}, and double-scaled versions of this model have been studied in \cite{Berkooz:2024evs,Berkooz:2024ofm,Almheiri:2024xtw}. A recent extension with multiple commuting DSSYK models, which collectively can describe a chaotic model, was developed in \cite{Gao:2024lem}.

Meanwhile, the integrable part of (\ref{eq:chaos-integrability interpolating Hamiltonian}) is constructed by combining $N$ Majorana fermions into Cartan generators of SO($N$), namely the fermion bilinears $\rmi Q_k=\psi_{2k-1}\psi_{2k}$, leading to \cite{Berkooz:2024evs,Berkooz:2024ofm,Gao:2023gta}
    \begin{equation}\label{eq:Hamiltonian integrable}
        {H}^{(p/2)}_{\rm integrable}=\sum_{\tilde{I}}B_{\tilde{I}}Q_{\tilde{I}}\,,
    \end{equation}
    where $\tilde{I}$ indicates $1\leq i_1\leq \dots\leq i_{p/2}\leq N/2$, $B_{\tilde{I}}\equiv B_{i_1,\dots,~i_{p/2}}$, $Q_{\tilde{I}}\equiv Q_{i_1}\dots Q_{i_{p/2}}$, where we take $p$ to be even, and we choose a Gaussian distribution of the kind 
    \begin{equation}
        \expval{B_{\tilde I}}=0\,,\quad
        \expval{B_{\tilde I}B_{\tilde I'}}=\delta_{\tilde I\tilde I'}\mathcal{J}^2
        \binom{N/2}{p/2}^{-1}\,,
    \end{equation}
where we have chosen the overall normalization to involve the same $\mathcal{J}$ parameter as in the DSSYK for convenience.

We will consider the double-scaling limit as in the previous section. The respective auxiliary system spanned by orthonormal chord number states $\qty{\ket{z}}$, and the doubled-scaled Hamiltonian is denoted $\hH_2$ in the auxiliary Hilbert space.

Let the penalty factors between different chord intersections be denoted as follows:
\begin{subequations}
\begin{align}
 {H}_{\rm SYK}\cap{H}_{\rm SYK}:~q_{nn}\,,\quad {H}_{\rm SYK}\cap {H}_{\rm integrable}:~q_{nz}\,,\quad{H}_{\rm integrable}\cap{H}_{\rm integrable}:~q_{zz}\,.
\end{align}
\end{subequations}
The moments of the ensemble averaged Hamiltonian (\ref{eq:chaos-integrability interpolating Hamiltonian}) can be expressed in terms of chord diagrams as
\begin{equation}\label{eq:chord tech}
    \bra{0}\hH^k\ket{0}=\sum_{\text{chord diagrams}}\nu^n\kappa^z q_{nn}^{\#\qty({H}_{\rm SYK}\cap{H}_{\rm SYK})} q_{nz}^{\#({H}_{\rm SYK}\cap{H}_{\rm integrable})} q_{zz}^{\#({H}_{\rm integrable}\cap{H}_{\rm integrable})}\,,
\end{equation}
where $k=n+z$; $\ket{0}$ is the zero-chord state, $q_{ij}=\rme^{-\lambda_{ij}}$ are the intersection weights (here $i,j=\qty{n,z}$) are associated with $\hH_1$ and $\hH_2$; and \# denotes number of intersections. In particular, the \emph{integrable-chaotic} case we have that $q_{nn}=q_{nz}=q$, $q_{zz}=1$ \cite{Berkooz:2024ofm}.

The chord diagrams can then be used to define an auxiliary Hilbert space for the BBJM model \cite{Berkooz:2024evs,Berkooz:2024ofm} with $k$ chords
\begin{equation}\label{eq:meaning r}
    \mathcal{H}_k={\rm span}\qty{\ket{n,~z;~\vec{r}}\,,~\text{s.t.}\quad\vec{r}=\qty{0,1}\,,\quad\sum_{i=1}^{k=n+z}r_i=n}\,,
\end{equation}
with $\dim(\mathcal{H}_k)=2^k$, and we define a dual basis $\llangle{n,~z;~\vec{r}}|$ \cite{Berkooz:2024ofm}
\begin{equation}\label{eq:matrix element}
    \llangle{n,~z;~\vec{r}}\ket{n',~z';~\vec{r}'}=\delta_{nn'}\delta_{zz'}\delta_{\vec{r}\vec{r}'}\,.
\end{equation}
In the integrable-chaotic case, the transfer matrix of the ensemble-averaged theory, which plays the role of an effective Hamiltonian, can be written in terms of a q-deformed and a standard harmonic-oscillator algebra acting on the auxiliary Hilbert space of the full system,
\begin{equation}
    \label{eq:transfer_mat}
    \hH=\nu(\ha_n+\ha_n^\dagger)+\kappa(\ha_z+\ha_z^\dagger)\,,
\end{equation}
where $\ha_{n/z}^\dagger$ and $\ha_{n/z}$ are the creation and annihilation operators acting on \eqref{eq:meaning r} as 
\begin{align}
    \ha_{n}\ket{n,z,\vec{r}}&=\sqrt{[n]_q}\ket{n-1,z,\vec{r}_1}\,,\quad &&\ha_{n}^\dagger\ket{n,z,\vec{r}}=\sqrt{[n+1]_q}\ket{n+1,z,\vec{r}_2}~ ,\\
    \ha_{z}\ket{n,z,\vec{r}}&=\sqrt{z}\ket{n,z-1,\vec{r}_3}\,,\quad &&\ha_{z}^\dagger\ket{n,z,\vec{r}}=\sqrt{z+1}\ket{n,z+1,\vec{r}_4}~ ,
\end{align} 
where $[n]_q:=\frac{1-q^n}{1-q}$, and $\vec{r}_{1\leq i\leq4}$ represent different vectors determining the state \eqref{eq:meaning r}. The total chord number operator is defined \cite{Berkooz:2024ofm}
\begin{eqnarray}\label{eq:chordnumber}
    \frac{1}{\lambda}\hat{l}\ket{n,z,\vec{r}}=(n+z)\ket{n,z,\vec{r}}\,.
\end{eqnarray}
The two-point function of a normalized chaotic matter operators $O$ (defined in \eqref{eq:chord matter operator}) with penalty factor $\rme^{-\bar \lambda}$ with the chaotic chords and integral chords could be computed by chord diagram with\footnote{This follows similarly to (B.30) in \cite{Berkooz:2024ofm} using the corresponding intersection factors between chaotic and integrable chords.},
\begin{equation}
    \begin{aligned}
    G_\Delta(\tau_1,\tau_2)
    &\,\equiv\avg{\Tr[\rme^{-(\beta+\tau_1)\hH}O\rme^{-(\tau_1-\tau_2)\hH}Oe^{-\tau_2 \hH}]/\Tr[1]}\\
    &\,={\bra{0}\rme^{-(\beta-\tau_2+\tau_1)\hH}\rme^{{{-}}\Delta \hat{l}}\rme^{-(\tau_2-\tau_1)\hH}\ket{0}}\,,
\end{aligned}
\end{equation}
where $\Delta=\bar\lambda/\lambda$. Due to the time-translational invariance, we have $G_\Delta(\tau_1,\tau_2)=G_\Delta(\tau_{12})$ with $\tau_{12} \equiv\tau_1-\tau_2$.

\subsection{Effective action and phase transition}\label{ssec:correlation}

{We will first review the path integral for the variables $n$ and $z$
\begin{align}
    \bra0e^{-\beta \hat H}\ket0 \propto Z_\beta=\int DnDz e^{-\frac1\lambda S}\,,
\end{align}
where $n(\tau_1,\tau_2)/\lambda$ and $z(\tau_1,\tau_2)/\lambda$ represent the densities of chaotic and integrable chords connecting points $\tau_1$ and $\tau_2$ in the thermal circle and the effective action is
\begin{align}\label{eq:effective_action_nz}
    S=&\,\frac14 \int_0^\beta d\tau_1\int_0^\beta d\tau_2 \int_{\tau_1}^{\tau_2}d\tau_3 \int_{\tau_2}^{\tau_1} d\tau_4 \kd{n(\tau_1,\tau_2)n(\tau_3,\tau_4)+2n(\tau_1,\tau_2)z(\tau_3,\tau_4)}\\
    &\,+ \frac12 \int_0^\beta d\tau_1 \int_0^\beta d\tau_2 \kd{n(\tau_1,\tau_2)\kc{\log\frac{n(\tau_1,\tau_2)}{\nu^2\J^2}-1} + z(\tau_1,\tau_2)\kc{\log\frac{z(\tau_1,\tau_2)}{\kappa^2\J^2}-1}} \nn\,,
\end{align}
where, in the coarse graining scheme for chords, the steps including (a) Generate, (b) Split, and (c) Reorder, contribute to the second line and the step (d) Cross, contributes to the first line, where we use the terminology in \cite{Berkooz:2024evs}.}

{We consider the classical limit $\lambda\to0$ and derive the Liouville equation from the variation principle  
\begin{align}\label{eq:Liouville}
\begin{split}
    &\partial_1\partial_2 g_n(\tau_1,\tau_2)+2\nu^2\J^2 e^{g_n(\tau_1,\tau_2)+g_z(\tau_1,\tau_2)}=0\,,\\
    &\partial_1\partial_2g_z(\tau_1,\tau_2)+2\kappa^2\J^2 e^{g_n(\tau_1,\tau_2)}=0
\end{split}
\end{align}
where we introduced the new variables
\begin{align}\label{eq:explicit gnz}
    g_n(\tau_1,\tau_2)=-\int_{\tau_1}^{\tau_2} d\tau_3 \int_{\tau_2}^{\tau_1} d\tau_4 n(\tau_3,\tau_4),\quad g_z(\tau_1,\tau_2)=-\int_{\tau_1}^{\tau_2} d\tau_3 \int_{\tau_2}^{\tau_1} d\tau_4 z(\tau_3,\tau_4),
\end{align}
where $g_n(\tau_1,\tau_2)/\lambda$ and $g_z(\tau_1,\tau_2)/\lambda$ respectively describe the numbers of chaotic and integrable chords going through a line connecting $\tau_1$ and $\tau_2$. 
It is convenient to define their sum 
\begin{align}\label{eq:l}
    l(\tau_1,\tau_2)=-\qty(g_n(\tau_1,\tau_2)+g_z(\tau_1,\tau_2))\,,
\end{align}
which is the value of chords number operator $\hat l$ sandwiched by two states, namely 
\begin{align}    
l(\tau_1,\tau_2)=\frac{\bra{0}\rme^{-(\tau_2-\tau_1)\hH}\hat{l}\rme^{-(\tau_1-\tau_2+\beta)\hH}\ket{0}}{\bra{0}e^{-\beta \hH}\ket{0}}.
\end{align}

In the classical limit $\lambda\to0$,  the two-point function $G_\Delta$ of chaotic matter operators $O$ \eqref{eq:chord matter operator} are related to the number of chords of both types that cross the chaotic matter chord \cite{Berkooz:2024ofm}
\begin{align}\label{eq:autocorre}
    \lim_{\lambda\rightarrow0}G_\Delta(\tau_1,\tau_2)&=\rme^{-\Delta l(\tau_1,\tau_2)}=\rme^{\Delta(g_n(\tau_1,\tau_2)+g_z(\tau_1,\tau_2))}\,,
\end{align}
where $g_{n/z}$ and $l$ are defined in \eqref{eq:explicit gnz}\eqref{eq:l} and could be computed by solving \eqref{eq:Liouville}. 

Due to time-translational invariance, we have $l(\tau_1,\tau_2)=l(\tau)$ and $g_{n/z}(\tau_1,\tau_2)=g_{n/z}(\tau)$, where $\tau\equiv \tau_1-\tau_2$. The double-time Liouville equations \eqref{eq:Liouville} reduce to the equations in terms of $\tau$,
\begin{align}\label{eq:euclidean EOM}
    \partial_\tau^2 g_n(\tau)=2\J^2\nu^2e^{g_n(\tau)+g_z(\tau)}\,,\quad\partial_\tau^2 g_z(\tau)=2\J^2\kappa^2e^{g_n(\tau)}.
\end{align}
Since no chord going through the connection line when $\tau_1\to\tau_2$, the imaginary-time boundary conditions are \cite{Berkooz:2024ofm}
\begin{align}\label{eq:Eulcidean IC}
    g_{n/z}(0)=g_{n/z}(\beta)=0\,,
\end{align}
in agreement with the condition of the Euclidean two-point functions $G_\Delta(0)=G_\Delta(\beta)=1$ via  \eqref{eq:autocorre}.

Plugging back the solutions $g_{n/z}$ in the action \eqref{eq:effective_action_nz} gives the on-shell action \begin{eqnarray}\label{eq:S on}
    S_{\rm on-shell}=\frac{1}{4}\iint_0^\beta\rmd\tau_1\rmd\tau_2\kd{\nu ^2 (g_n-2) e^{g_n+g_z}+2 \kappa ^2 (g_n-1)e^{g_n}}\,.
\end{eqnarray}
There are three different saddle point solutions to the equations of motion \eqref{eq:Liouville} depending on $\kappa$, corresponding to the quasi-integrable, chaotic and a subleading saddle point. These three solutions lead to a thermodynamic phase transition in terms of the free-energy in the saddle point approximation \cite{Berkooz:2024ofm}.

We  numerically solve the equation  and evaluate the on-shell action for various parameters $\beta$ and $\kappa$. 
As shown in Fig.\,\ref{fig:action}, the kink of the free energy given by the on-shell action indicates the phase transition.
\begin{figure}
    \centering
    \includegraphics[width=0.75\textwidth]{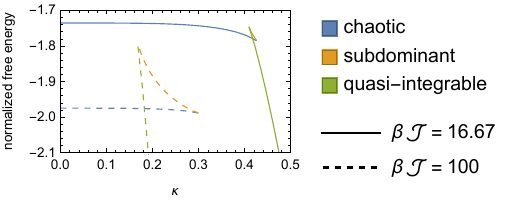}
    \caption{Normalized free energy $S_\text{on-shell}/(\lambda\beta\J)$ as a function of the parameter $\kappa$ at $\beta\J=16.67,\,100$. The curves indicate the phase transition points at $\kappa= 0.417,\,0.18$.}
    \label{fig:action}
\end{figure}

Next, we discuss perturbative solutions about these saddle point solutions, which were previously found in \cite{Berkooz:2024ofm}.

\paragraph{Chaotic saddle}
For $\kappa\ll1$, we have as the leading order in $\kappa$ solution to \eqref{eq:Liouville} and \eqref{eq:Eulcidean IC}:
\begin{align}\label{eq:euclidean chaotic saddle}
        l(\tau) ={{-}}2\log\left[\frac{\cos\left(\frac{\pi v}{2}\right)}{\cos\left[\pi v\left(\frac{1}{2} - \frac{\tau}{\beta}\right)\right]}\right]+\mathcal{O}(\kappa^4)\,,\quad g_z(\tau)=2\kappa^2\log\left[\frac{\cos\left(\frac{\pi v}{2}\right)}{\cos\left[\pi v\left(\frac{1}{2} - \frac{\tau}{\beta}\right)\right]}\right]+\mathcal{O}(\kappa^4)\,,
\end{align} 
where we have defined
\begin{equation}\label{eq:def v}
    {\beta \mathcal{J} \equiv \frac{\pi v}{\cos\left(\frac{\pi v}{2}\right)}\,.}
\end{equation}
Thus, at the leading order, we have $g_z = 0$, and $\ell$ is simply the single chord solution with (dimensionless) temperature $\beta \mathcal{J}$.

\paragraph{Integrable saddle}
Similarly, considering $\nu\ll1$ and low temperatures, $\beta \mathcal{J} \gg 1$, we have as the leading order solutions of  \eqref{eq:Liouville} and \eqref{eq:Eulcidean IC} 
\begin{equation}\label{eq:int saddles}
   l(\tau) = {{-}}\left( \mJ\kappa\right)^{2}\tau\left(\tau - \beta\right)+O\left(\frac{(\beta \mJ)^0}{\kappa^2}\right) \, , \quad g_n(\tau)=O\left(\frac{\nu^2}{\kappa^4(\beta \mJ)^2}\right)\,.
\end{equation}
So at low temperatures, it is consistent to consider only $l(\tau)$ in \eqref{eq:euclidean chaotic saddle}  \eqref{eq:int saddles} to evaluate the autocorrelation function which will be defined in Sec.\,\ref{sec:Krylov operators}. We use these saddle point solutions to evaluate the Krylov operator complexity in the next section.

\subsection{Evolution of chord numbers}\label{ssec:evol phases}
We now evaluate the chord number in \eqref{eq:l} in Lorentzian signature by implementing a (Hartle-Hawking \cite{Hartle:1983ai}) analytic continuation at finite temperature: $\tau\rightarrow \beta/2+\rmi t$ in the differential equations (\ref{eq:Liouville}),
\begin{subequations}\label{eq:all eqs}
    \begin{align}
    \partial_{t}^2\tilde{g}_{n}(t) &=-2\mathcal{J}^2 \nu^2 \rme^{\tilde{g}_{n}(t)+\tilde{g}_{z}(t)},\label{eq:g_n}\\
    \partial_{t}^2\tilde{g}_{z}(t)& =-2\mathcal{J}^2 \kappa^2 \rme^{\tilde{g}_{n}(t)},\label{eq:g_z}
\end{align}
\end{subequations}
where we denote 
\begin{eqnarray}\label{eq:rotation g}
\tilde{g}_{n/z}(t)\equiv g_{n/z}(\tau=\beta/2+\rmi t)\,,
\end{eqnarray}
while the boundary conditions (\ref{eq:Eulcidean IC}) are inherently Euclidean. The initial conditions for real time evolution is given by the solution from imaginary time and the time-reversal symmetry 
\begin{align}\label{eq:IV}
    \tilde g_{n/z}(0)=g_{n/z}(\beta/2),\quad \tilde g_{n/z}'(0)=0\,.
\end{align}
where the time-reversal symmetry imply that $\tilde g_{n/z}(t)=\tilde g_{n/z}(-t)$. Similar to \eqref{eq:l}, we define the rescaled total chord number on real time 
\begin{eqnarray}\label{eq:def l(t)}
l(t)\equiv-\qty( \tilde{g}_{n}(t)+\tilde{g}_{z}(t))\,.
\end{eqnarray}

\begin{figure}
    \centering
    \includegraphics[width=0.65\linewidth]{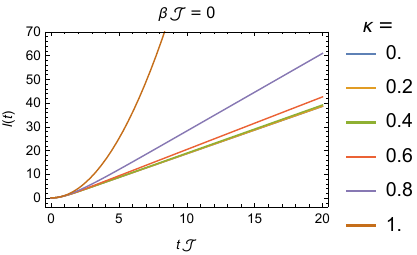}
    \caption{Evolution of $l(t)$ \eqref{eq:def l(t)} along real time, at $\beta\J=0$ and $\kappa=0,0.2,0.4,0.6,0.8,1.0$. The integrable system limit ($\kappa=1$) displays the expected parabolic growth; while the chaotic one ($\kappa=0$) displays late time linear growth, as seen in \eqref{eq:int saddles} and \eqref{eq:euclidean chaotic saddle} with $\tau=\rmi t$ ($\beta=0$) respectively. Note that the blue, amber and green lines nearly overlap with each other.}
    \label{fig:time dependence g}
\end{figure}
\begin{figure}
    \centering   \includegraphics[width=0.85\linewidth]{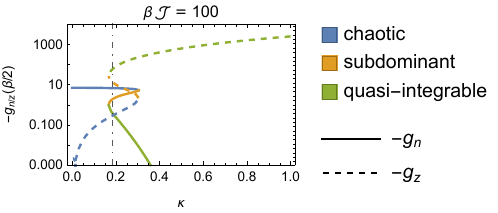}
    \caption{$-g_n(\beta/2)$ and $-g_z(\beta/2)$ \eqref{eq:explicit gnz} as functions of $\kappa$ at $\beta\J=100$ for all three branches of saddles. The dot-dashed line denotes the phase transition point $\kappa=0.18$. The values of $-g_n(\beta/2)$ and $-g_z(\beta/2)$ in the dominated saddle exhibit discontinuous jumps when crossing the first-order transition.}
    \label{fig:gnzkappa}
\end{figure}
\begin{figure}
    \centering
    \includegraphics[width=0.9\linewidth]{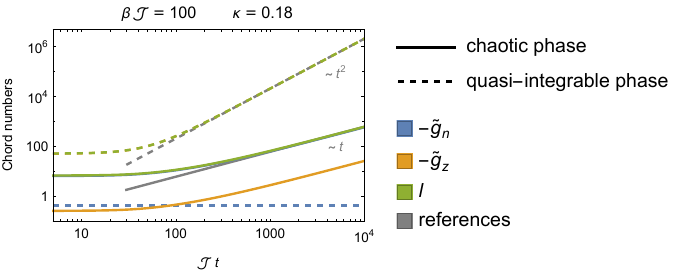}
    \caption{Real time evolution of $l(t)=-\tilde g_n(t)-\tilde g_z(t)$ in different phases near the phase transition line at $\beta\J=100$ and $\kappa=0.18$. 
    The solid blue curve is covered by the green curve and the orange dashed curve is covered by the green dashed curve.}
    \label{fig:RealtimeEvoPhases}
\end{figure}

We expect that, at fixed $\beta \J$, the growth rate of the total chord number $l(t)$ increases with increasing $\kappa$. As a consequence, the growth in the chaotic phase is slower than that in the quasi-integrable phase. This behavior can be understood from the dominant contributions to the chord number. For small $\kappa$, $l(t)$ is dominated by $\tilde g_n(t)$, whose evolution is governed by \eqref{eq:g_n}. Its growth is strongly constrained by the $\rme^{\tilde g_n(t)}$ damping and therefore remains limited. In contrast, for large $\kappa$, $l(t)$ is dominated by $\tilde g_z(t)$, whose dynamics is governed by \eqref{eq:g_z}. In this regime, the $\rme^{\tilde g_n(t)}$ damping becomes negligible, allowing the chord number to grow more rapidly. This contrast is particularly pronounced at large $\beta \J$, where a strong hierarchy develops between $\tilde g_n(t)$ and $\tilde g_z(t)$. Consequently, at large $\beta \J$, the chord number grows much faster in the quasi-integrable phase than in the chaotic phase.

To investigate the growth of the chord number for general $\beta \J$, we first numerically solve the imaginary-time Liouville equation \eqref{eq:euclidean EOM} with the boundary condition \eqref{eq:Eulcidean IC}, and then numerically solve the real-time Liouvillian equation \eqref{eq:all eqs} with the initial conditions \eqref{eq:IV}.

At $\beta \J = 0$, the system undergoes a crossover from the chaotic phase to the quasi-integrable phase as $\kappa$ increases. Correspondingly, the growth rate of $l(t)$ increases smoothly, and the growth behavior crosses over from linear to quadratic, as shown in Fig.\,\ref{fig:time dependence g}.

At $\beta \J = 100$, the imaginary-time solution exhibits a discontinuous jump at the first-order transition point $\kappa = 0.18$, as shown in Fig.\,\ref{fig:gnzkappa}. We then take the solutions on the two sides of the transition as initial conditions and compute their real-time evolutions, respectively. The results are shown in Fig.\,\ref{fig:RealtimeEvoPhases}. In addition to the difference in the magnitude of the chord number, we observe that the chord number grows linearly in the chaotic phase and quadratically in the quasi-integrable phase, indicating a discontinuous change in the growth behavior.

The crossover and the jump in the chord number and its growth behavior are consequences of the thermodynamic crossover and phase transition, respectively. However, the chord number itself does not directly characterize chaos or scrambling. To analyze these properties, one must instead investigate measures of scrambling, such as the Krylov operator complexity and the operator size, across the phase diagram, particularly in the vicinity of the transition line.

\section{Construction of the Krylov basis}\label{sec:krylov basis spread}

We now construct the Krylov basis of the Krylov state complexity with respect to the infinite-temperature TFD state and show there is a non-trivial limit in which we may identify the total chord number involving integrable and chaotic chords with the Krylov number operator. 
In Sec.~\ref{ssec:Krylov basis for all q} we derive the first few Krylov basis in the Lanczos algorithm for general intersection parameter $q_{nn}$, $q_{nz}$ and $q_{zz}$. 
In Sec.~\ref{ssec:special case} we specialize in the case $q_{nn}=q_{nz}=q_{zz}$ where we recover a closed form expression for the Krylov basis, Lanczos coefficients, and spread complexity\footnote{However, given that this case does not describe the integrable-chaotic transition, we will not discuss this part after this section.}.

\subsection{General Krylov basis}\label{ssec:Krylov basis for all q}
We consider the monic version of the Lanczos algorithm \cite{Muck:2022xfc} in which the diagonal coefficients $a_n$ do not appear, as the density of states in the partition function are energy eigenvalue-symmetric,
\begin{align}\label{eq:monic lanczos}
    \ket{P_{n+1}}=\hH\ket{P_{n}} - b_n^2\ket{P_{n-1}},\quad b_n^2=\frac{\avg{P_n|P_n}}{\avg{P_{n-1}|P_{n-1}}}
    =\frac{\bra{P_{n-1}}\hH\ket{P_n}}{\avg{P_{n-1}|P_{n-1}}}\,.
\end{align}
Based on this Krylov basis, the Krylov complexity operator is defined by
\begin{align}\label{eq:ell op def}
    \frac{\hat{l}}{\lambda}=\sum_{k=0}^\infty k\frac{\ket{P_k}\bra{P_k}}{\avg{P_k|P_k}}\,.
\end{align}
In the following, we show that when all the Hamiltonian intersections are equal to each other, the Krylov basis $\ket{P_k}$ for the BBJM model can be expanded by the chords states $\ket{n,z,\vec r}$ in (\ref{eq:meaning r}) with $k=n+z$ as the total chord number. Starting from
\begin{eqnarray}
 \ket{P_0}=\ket{0,0,\vec{0}},\quad b_0^2=0\,,   
\end{eqnarray}
the Lanczos algorithm leads to,
\begin{subequations}
    \begin{align}
\ket{P_1}&=\nu\ket{1,0,\vec{r}_1}+\kappa\ket{0,1,\vec{r}_2},\quad b_1^2=\nu^2+\kappa^2=1,\\
\ket{P_2}&=\nu^2\ket{2,0,\vec{r}_3}+\nu\kappa\ket{1,1,\vec{r}_4=(1,0)}+\nu\kappa\ket{1,1,\vec{r}_5=(0,1)}+\kappa^2\ket{0,2,\vec{r}_6}~,\label{eq:P2 explicit}\\ 
b_2^2&=\nu^4(1+q_{nn})+2\nu^2\kappa^2(1+q_{nz})+\kappa^4(1+q_{zz})\label{eq:b_2}
\end{align}
\end{subequations}
where the notation (0,1,\dots) indicates how to build the corresponding $\vec{r}_i$ above (see \eqref{eq:meaning r}). Similar expressions follow for $\ket{P_3}$; however, by computing the overlap between the previous states we obtain
\begin{subequations}\label{eq:subeqs}
    \begin{align}
    \llangle 1,0,\vec{r}_1| \hH\ket{P_2}&=\nu \kd{\nu^2(1+q_{nn})+\kappa^2(1+q_{nz})}, \\
\llangle 0,1,\vec{r}_2| \hH\ket{P_2}&=\kappa \kd{\nu^2(1+q_{nz})+\kappa^2(1+q_{zz})},\\
\llangle 1,0,\vec{r}_1\ket{P_3}&=\nu \kd{\nu^2(1+q_{nn})+\kappa^2(1+q_{nz})-b_2^2}\\
&=\nu\kappa^2\kd{\nu^2(q_{nn}-q_{nz})+\kappa^2(q_{nz}-q_{zz})},\nonumber\\
\llangle 0,1,\vec{r}_2\ket{P_3}&=\kappa \kd{\nu^2(1+q_{nz})+\kappa^2(1+q_{zz})-b_2^2}\\
&=\kappa\nu^2\kd{\nu^2(q_{nz}-q_{nn})+\kappa^2(q_{zz}-q_{nz})}\,.\nonumber
\end{align}
\end{subequations}
This implies that we can express
\begin{eqnarray}\label{eq:P3 case}
    \ket{P_3}=\sum_{n+z=3,\vec r} \nu^n\kappa^z\ket{n,z,\vec r}+\ket{\text{1-chord}}\,.
\end{eqnarray}
where $\ket{\text{1-chord}}$ denotes a linear combination of $\ket{1,0,\vec{r}_1}$ and $\ket{0,1,\vec{r}_2}$. Note, \eqref{eq:P3 case} implies that the third element in the Krylov basis corresponds to a one-chord state with and a state with fixed total chaotic and integrable chord number equal to the ``position'' in the Krylov chain. Iterating the above result for the other elements in the Krylov basis, we expect that $\llangle n,z,\vec r\ket{P_k}$ for $n+z\neq k$ is linear in $q_{nn}-q_{nz}$ and $q_{zz}-q_{nz}$. We stress that one can carry out the algorithm above to build the Krylov basis at arbitrary order $\ket{P_n}~\forall\kappa\in[0,1]$, but it is technically intricate due to the evaluations involved. However, we expect that the algorithm to higher order will generically lead to an expression where $\ket{P_{n}}$ is manifestly linearly dependent of lower-order chord states (i.e. $\ket{\rm 1-chord}$, \dots $\ket{(m<n)-\rm chord}$).

In addition, we show in App.\,\ref{app:alt exp} that $\qty{\ket{n,z;\vec{r}}}$ is not generically a Krylov basis. Nevertheless, we find the full Krylov basis in this system for a special case of the Hamiltonian intersections below.

\subsection{Generalization}

In the following, we focus on the integrable-chaotic transition, for which $q_{nn}=q_{nz}=q$, and $q_{zz}=1$.

Based on our finding for the $k=3$ Krylov basis \eqref{eq:subeqs}, we assume that at least for a given order $k\geq3$ the Krylov basis takes the form 
\begin{eqnarray}\label{eq:ansatz k}
    \ket{P_k}=\sum_{n+z=k,\vec{r}}\nu^n\kappa^z\ket{n,z,\vec{r}}+\ket{\xi_k}~,
\end{eqnarray}
where $\avg{\xi_k|\xi_k}\sim\mathcal{O}(\lambda^2)$.

We will show by induction that the form of the Krylov basis $\ket{P_k}$ holds for arbitrary higher orders. For this, we search for the next element in the Krylov basis
\begin{eqnarray}\label{eq:Pk+1}
    \ket{P_{k+1}}=\hH\ket{P_k}-b^2_k\ket{P_{k-1}}~,
\end{eqnarray}
where $\hH$ is the total chord Hamiltonian \eqref{eq:transfer_mat}. This gives
\begin{eqnarray}\label{eq:Pk extra}
    \begin{aligned}
        \hH\ket{P_k}=\sum_{n+z=k}\nu^{n}\kappa^z\Biggl(&\nu\qty(\sum_{\vec{r}'_1}\ket{n+1,z,\vec{r}'_1}+[k]_q\sum_{\vec{r}'_2}\ket{n-1,z,\vec{r}'_2})\\
        &+\kappa\qty(\sum_{\vec{r}'_3}\ket{n,z+1,\vec{r}'_3}+k\sum_{\vec{r}'_4}\ket{n,z-1,\vec{r}'_4})\Biggr)+O(\lambda)~,
    \end{aligned}
\end{eqnarray}
where $\vec{r}'_{1\leq i\leq 4}$ represents the vector that specifies each of the contributions with $k+1$ and $k-1$ total number of chords. Thus, \eqref{eq:Pk+1} gives
\begin{equation}
    \ket{P_{k+1}}=\sum_{n+z={k+1}}\sum_{\vec{r}}\nu^n\kappa^z\ket{n,z,\vec{r}}+\ket{\xi_{k+1}}~,
\end{equation}
where $\ket{\xi_{k+1}}$ are the states with $k-1$ total chord number in \eqref{eq:Pk+1} and \eqref{eq:Pk extra}. We search to evaluate the norm of $\ket{\xi_{k+1}}$ in the $\lambda\rightarrow0$ limit. To this effect, we compute the following overlaps,
\begin{equation}
\begin{aligned}
    \llangle n-1,z,\vec{r}|\hH\ket{P_k}&=(\nu^2+\kappa^2)[k]_q\nu^{n-1}\kappa^z~,\\
    \llangle n,z-1,\vec{r}|\hH\ket{P_k}&=(\nu^2[k]_q+\kappa^2k)\nu^n\kappa^{z-1}~.
\end{aligned}
\end{equation}
where the double bracket is given by \eqref{eq:matrix element}.
Next, we use the explicit form of the Krylov basis \eqref{eq:ansatz k} to compute the Lanczos coefficient \eqref{eq:monic lanczos}
\begin{equation}
\begin{aligned}
    b_k^2&=\frac{\langle P_{k-1}|\hH\ket{P_k}}{\avg{P_{k-1}|P_{k-1}}}=(\nu^4+2\nu^2\kappa^2)[k]_q+\kappa^4 k~.
\end{aligned}
\end{equation}
Then, we can compute the norm of the additional term
\begin{eqnarray}
&\begin{aligned}
    \llangle{n-1,z,\vec{r}}\ket{P_{k+1}}=&\llangle{n-1,z,\vec{r}}|\hH\ket{P_k}-b_{k}^2\llangle{n-1,z,\vec{r}}\ket{P_{k-1}}\\
    =&\nu^{n-1}\kappa^z\qty([k]_q-b_{k}^2)=\nu^{n-1}\kappa^{z+4}([k]_q-k)~,
\end{aligned}\\
&\begin{aligned}
    \llangle{n,z-1,\vec{r}}\ket{P_{k+1}}=&\llangle{n,z-1,\vec{r}}|\hH\ket{P_k}-b_{k}^2\llangle{n,z-1,\vec{r}}\ket{P_{k-1}}\\
    =&\nu^{n}\kappa^{z-1}\qty(\nu^2[k]_q+\kappa^2k-b_{k}^2)=\nu^{n+2}\kappa^{z+1}(k-[k]_q)~.
\end{aligned}
\end{eqnarray}
Then, in the semiclassical limit
\begin{eqnarray}\label{eq:new Pk+1}
    \llangle{n-1,z,\vec{r}}\ket{P_{k+1}}=O(\lambda)~,\quad \llangle{n,z-1,\vec{r}}\ket{P_{k+1}}=O(\lambda),
\end{eqnarray}
which implies that $\avg{\xi_{k+1}|\xi_{k+1}}\sim\mathcal{O}(\lambda^2)$. Thus, Krylov spread complexity is the expectation value of the total chord number in the $\lambda\rightarrow0$ limit. This implies that our results in Sec.~\ref{ssec:evol phases} can be interpreted as Krylov spread complexity.

\subsection{Special case: \texorpdfstring{$q_{nn}=q_{nz}=q_{zz}$}{}}\label{ssec:special case}
When $q_{nn}=q_{nz}=q_{zz}$, we can think of the total Hamiltonian $H$ (\ref{eq:chaos-integrability interpolating Hamiltonian}) as a superposition of $\nu H_1$ chord and $\kappa H_2$ chord with penalty factor $q_{nn}=q_{nz}=q_{zz}$. Then the $H$-chord state is the Krylov basis for $\ket{P_0}=\ket{0}$,
\begin{align}\label{eq:KrylovIdenticalChord}
    \ket{P_k}=\sum_{n+z=k,\vec r} \nu^n\kappa^z\ket{n,z,\vec r} \quad (q_{nn}=q_{nz}=q_{zz}=q)\,,
\end{align}
which implies
\begin{align}\label{eq:deviations}
    b_n^2=\frac{1-q^n}{1-q},\quad \mu_{2n}=\prod_{m=1}^n b_m^2=\frac{(q;q)_n}{(1-q)^n} \,,\quad n\geq1\,. 
\end{align}
Therefore, in this case the state Krylov complexity is the expectation value of the total chord number operator \eqref{eq:ell op def}. Given that the total chord number corresponds to the position operator in the Krylov chain, the corresponding Krylov complexity when $q_{nz}=q_{zz}=q_{nn}=q$ is the same as for the DSSYK model, namely (\ref{eq:sonner}).

Note that by the choice $q_{nz}=q_{zz}=q_{nn}=q$ the configuration essentially corresponds to a system of two decoupled DSSYK models of different colors.\footnote{DSSYK systems with several colors have attracted interest in the literature \cite{Berkooz:2024ifu,Gao:2024lem} motivated by experimental progress in quantum teleportation protocols in quantum computing \cite{Jafferis:2022crx}.} However, this system does not describe the BBJM model. Instead, the spread complexity follows just as in the DSSYK model. For completeness, we analyze deviations from the Krylov basis with $q_{nz}=q_{zz}=q_{nn}$ in App.\,\ref{ssec:deviations krylov basis}.

To summarize, the main outcome of this section is that the expectation value of the total chord number operator in the Hartle-Hawking state of the integrable-chaotic system \eqref{eq:chaos-integrability interpolating Hamiltonian} in the semiclassical limit is equal to Krylov complexity with the same Hartle-Hawking state as a reference state. Moreover, we found a specific regime, when all the penalty factors are equal one another, $q_{nn}=q_{zz}=q_{nz}$, where the same correspondence between total chord number and Krylov state complexity holds.}

\section{Krylov operator complexity across the phase transition}\label{sec:Krylov operators}

We now derive the Krylov operator complexity to diagnose the scrambling.  
In Sec.\,\ref{eq:moment method} we apply our results to compute Krylov operator complexity based on the moment method, where the normalized autocorrelation functions determine Krylov complexity. In Sec.\,\ref{ssec:perturbative} we evaluate Krylov operator complexity perturbatively based on the corresponding two-point functions.

\subsection{Krylov operator complexity}\label{eq:moment method}

As  the initial operator in the Krylov approach, we consider  a light operator $O$ with $\Delta\to0$ in the center of a half thermal circle, namely ${O}_\beta=e^{-\beta \hH/4}Oe^{-\beta \hH/4}$. 
In the Liouville operator Hilbert space, we denote by $\ket{O_\beta}$ the state corresponding to the operator $O_\beta$, with inner product
$\langle A|B\rangle ~\equiv \tr\big(A^\dagger B\big)~$,
so that $\langle O_\beta|O_\beta\rangle ~=\tr(O_\beta^\dagger O_\beta)$. Its normalized autocorrelation function is 
\begin{align}
    G(t)
    =\frac{\avg{O_\beta|e^{i\hL t}|O_\beta}}{\avg{O_\beta|O_\beta}}
    =1-\Delta ~(l(t)-l(0))+\cdots
    =e^{-\Delta (l(t)-l(0))}+\cdots
\end{align}
where the ellipsis represents higher order terms in the small $\Delta$ expansion. We construct the moments as
\begin{align}
    \mu_{n}=\frac{\avg{O_\beta|\hL^n|O_\beta}}{\avg{O_\beta|O_\beta}}=(-i)^nG^{(n)}(0)
    =\begin{cases}
        1,& n=0,\\
        -(-i)^n\Delta~{l^{(n)}(t=0)}+\cdots ,&  n\geq1.
    \end{cases}
\end{align}
Based on the Liouville equation in real time \eqref{eq:all eqs} we now calculate all the derivatives $\tilde g_{n/z}^{(n)}(0)$ from the iterative equations at $t=0$ where the superindex ${}^{(n)}$ indicates the $n$-th derivative with respect to $t$, 
\begin{align}
    &\tilde g_n^{(n+2)}(0)=-2\J^2\nu^2 B_n(-l^{(1)}(0),-l^{(2)}(0),\cdots,-l^{(n)}(0)) e^{-l(0)},\\ 
    &\tilde g_z^{(n+2)}(0)=-2\J^2\kappa^2 B_n(\tilde g_n^{(1)}(0),\tilde g_n^{(2)}(0),\cdots,\tilde g_n^{(n)}(0))(0)e^{\tilde g_n(0)},\\
    & l^{(n)}(0)=-\tilde g_n^{(n)}(0)-\tilde g_z^{(n)}(0), 
\end{align}
starting from the initial condition $\tilde g_{n/z}(0)=g_{n/z}(\beta/2),\,\tilde g_{n/z}'(0)=0$, where $B_n(x_1,x_2,\cdots,x_n)$ is the complete Bell polynomial. We carry this out numerically for general $\nu,\kappa$ and $\beta$. Since the differential equations and the initial conditions are invariant under time-reverse $t\to-t$, we have $\tilde g_{n/z}^{(n)}(0)=0$ for odd $n$, such that $\mu_n=0$ for odd $n$. We then calculate the moments $\mu_{2n}$ for even $n\leq n_c$ with $n_c$ an appropriate numerical truncation. In the $\Delta\to0$ limit, $\mu_n, \forall n\geq2$ is linear in $\Delta$. From the truncated moments and their Hankel matrix $H_n=\ke{\mu_{i+j}}_{0\leq i,j\leq n}$, we calculate the truncated Lanczos coefficients as
\begin{align}
    a_n=0,\quad b_1^2=\det H_1,\quad \quad b_n^2=\frac{\det H_n \det H_{n-2}}{(\det H_{n-1})^2}, \quad  n\leq n_c.
\end{align}
When $\kappa<1$, we find that $b_1\sim \sqrt{\Delta}$, whereas $b_n$ for $n\geq 2$ are of order $\Delta^0$. 
To compare the Lanczos coefficients and the Krylov operator complexities for different values of $\kappa$, 
we uniformly rescale $b_n\to \tilde b_n=\sqrt{2\Delta} b_n/b_1$ for all $n$ such that $\tilde b_1=\sqrt{2\Delta}$. 
Throughout the remainder of this subsection, we will focus on the rescaled Lanczos coefficients $\ke{\tilde b_n^2}$ and the dynamics generated by them.

Remarkably, the above algorithm shows that the rescaled Lanczos coefficients $\tilde b_n$ for $n\geq 2$ depend only on the parameter $\kappa$ and a single initial condition $\tilde g_z(0)$. 
More precisely,
\begin{align}\label{eq:rescaled_Lanczos}
    \tilde b_1^2=2\Delta,\quad 
    \tilde b_2^2=2-\frac{2 \kappa ^4}{(\tilde{\nu }^2+\kappa ^2)^2},\quad 
    \tilde b_3^2=6-\frac{2 \kappa ^4 (3 \tilde{\nu }^2+2 \kappa ^2)}{(\tilde{\nu }^2+\kappa ^2)^2 (\tilde{\nu }^2+2 \kappa ^2)},~~\cdots~~,\ \text{where}\  
    \tilde\nu=\nu e^{\tilde g_z(0)/2}
\end{align}
and $e^{\tilde g_z(0)/2}$ effectively renormalizes the coupling $\nu$ of the SYK Hamiltonian $H_1$ into a different value $\tilde\nu$.

We will discuss the hierarchy and behavior of the Lanczos coefficients. Recall the $\Delta\to0$ limit at first, we have $b_1\ll b_n$ for $n\geq2$ except $\nu=0$ exactly. On one side, when $\kappa\ll1$, which happens in the deep chaotic phase, we have 
\begin{align}\label{eq:Lanczos_chaotic}
    \tilde b_1=\sqrt{2\Delta},\quad \tilde b_n=\sqrt{n(n-1)}+\mathcal O\qty(\frac{\kappa^4}{\tilde\nu^4}),\quad n\geq2,\quad \frac{\kappa}{\tilde\nu}\ll 1.
\end{align}
On the other side, when $\tilde\nu\ll1$, which happens in the deep quasi-integrable phase, we consider the small $\tilde\nu/\kappa$ expansion and find 
\begin{align}\label{eq:Lanczos_integrable}
    \tilde b_1=\sqrt{2\Delta},\quad 
    \tilde b_2= \frac{2\tilde\nu}{\kappa}+\mathcal O\qty(\frac{\tilde\nu^3}{\kappa^3})\ll1,\quad 
    \tilde b_n\approx \sqrt{2(n-1)}+\mathcal O\qty(\frac{\tilde\nu^2}{\kappa^2}),\quad n\geq3, \quad \frac{\tilde\nu}{\kappa}\ll1,
\end{align}
where the behavior $\sqrt{n(n-1)}$ is given by fitting. It implies the hierarchy $\tilde b_1\ll \tilde b_2\ll \tilde b_n$ with $n\geq3$ in the quasi-integrable phase. 
The asymptotic behaviors are derived from the perturbative approach in the next subsection.

\begin{figure}
    \centering
    \includegraphics[width=\linewidth]{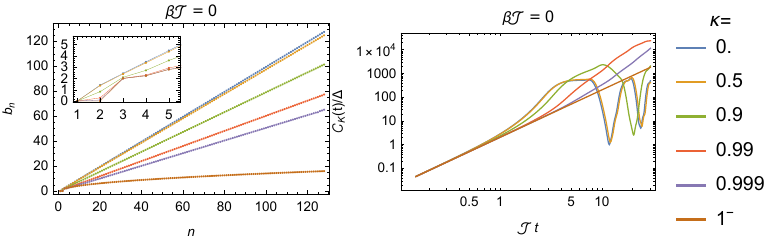}
    \caption{The Lanczos coefficients $\tilde b_n$ (left) and the Krylov complexity (right) for the matter chord operator defined in \eqref{eq:chord matter operator}  at $\beta\J=0$ for different values of $\kappa$ in the $\Delta\to0$ limit. The notation $1^-$ in the legend denotes a value infinitesimally smaller than $1$. The oscillations of the Krylov complexities are due to the truncation at $n_c=128$.}
    \label{fig:bnkckappa}
\end{figure}

\begin{figure}
    \centering
    \includegraphics[width=\linewidth]{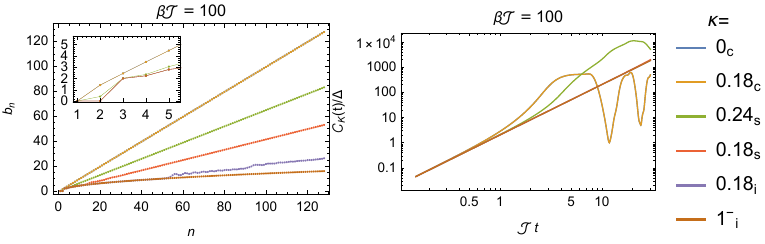}
    \caption{The Lanczos coefficients $\tilde b_n$ (left) and the Krylov complexity (right) for the matter chord operator of \eqref{eq:chord matter operator} at $\beta\J=100$ for different value of $\kappa$ in the $\Delta\to0$ limit. The subscripts in the values of $\kappa$ labels the corresponding saddle, where $c$ refers to chaotic saddle, $s$ refers to subdominant saddle and $i$ refers to quasi-integrable saddle. In both panels, the blue and amber dots(curves) are nearly coincident with each other. In the right panel, the red, purple and brown lines are coincident with each other.}
    \label{fig:bnkcbetakappa}
\end{figure}

We implement the above algorithm with exact numbers up to $n_c=60$ or high-precision numbers up to $n_c=128$.
At $\beta\J=0$, the Lanczos coefficients for some $\kappa$ are shown in the left panel of Fig.\,\ref{fig:bnkckappa}. Their behaviors continuously interpolate between the linear growth in \eqref{eq:Lanczos_chaotic} at small $\kappa$ and the square root growth in \eqref{eq:Lanczos_integrable} when $\kappa\to1$.  At $\beta\J=100$, with the transition value $\kappa=0.18$, the Lanczos coefficients are shown in the left panel of Fig.\,\ref{fig:bnkcbetakappa}. Following the saddle trajectory from the chaotic branch, through the subdominant branch, to the quasi-integrable branch, the Lanczos coefficients continuously transform from the $\propto n$ growth to the $\propto\sqrt{2n}$ growth. In the quasi-integrable phase, we observe $\tilde b_2 \ll \tilde b_n$ for $n\geq 3$, as expected.  
However, when tracking the dominated branches across the first-order phase transition, the Lanczos coefficients exhibit a discontinuous jump at the transition point, where $\tilde b_2$ drops sharply.

Then we can construct the tridiagonal form of the Liouvillian $\hL$ from the rescaled Lanczos coefficients $\ke{\tilde b_n}$, which is a  $(n_c+1)\times (n_c+1)$ matrix. Starting from the initial state $\varphi_n(0)=\delta_{n0}$, the Krylov wave function is
\begin{eqnarray}
    \varphi_m(t)=\sum_{n=0}^{n_c}\kc{e^{i\hL t}}_{mn}\varphi_n(0)\,.
\end{eqnarray}
This truncation remains valid for a finite time interval, during which 
$n_c\abs{\varphi_{n_c}(t)}^2$ remains negligible. 
Beyond this time regime, the reflections at the artificial boundary $n=n_c$ occur, 
which contaminate the subsequent evolution.

Finally, we calculate the Krylov complexity from 
\begin{eqnarray}
    \C_K(t)=\sum_{n=0}^{n_c}n\abs{\varphi_n(t)}^2\,.
\end{eqnarray}
Since $\tilde b_0\propto\sqrt\Delta$, we find that $\C_K(t)/\Delta$ converges to a finite value in the limit of $\Delta\to0$. 
At $\beta\J=0$, the Krylov complexity is shown in the right panel of Fig.\,\ref{fig:bnkckappa}. 
It displays a crossover from exponential growth to quadratic growth, with a \emph{decreasing} growth rate as $\kappa$ increases, 
indicating a crossover from the chaotic regime to the integrable regime. 
The oscillations visible in the Krylov complexity are artifacts of the truncation at $n_c$ and should be disregarded. 
At $\beta\J=100$, the Krylov complexity is shown in the right panel of Fig.\,\ref{fig:bnkcbetakappa}. Following the saddle trajectory through the chaotic--subdominant--quasi-integrable branches, the Krylov complexity exhibits a crossover from exponential growth to quadratic growth. However, when tracking the dominant branches across the first-order phase transition, the Krylov complexity exhibits a discontinuous jumps between the exponential growth and the quadratic growth at the transition point.

Finally, we observe that in certain cases -- such as the subdominant saddle at 
$\beta\J=100,\,\kappa=0.18$ shown in Fig.\,\ref{fig:bnkcbetakappa} -- the Lanczos coefficients $b_n$ 
exhibit global linear growth, while the Krylov operator complexity grows quadratically rather than exponentially. 
At first sight, this appears to contradict the general expectation of \cite{Parker:2018yvk}. 
The origin of this behavior lies in the hierarchy $b_2 \ll b_n$ for $n\geq 3$. 
This strong separation of scales causes $\varphi_1(t)$ to evolve much faster than 
$\varphi_n(t)$ for any $n\geq 2$, effectively reducing the dynamics to an approximately integrable scenario and leading to quadratic growth of the Krylov complexity.

\begin{figure}
    \centering
    \subfloat[]{\includegraphics[width=0.49\linewidth]{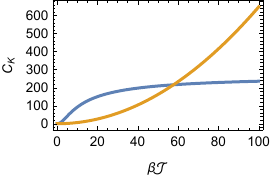}}\subfloat[]{\includegraphics[width=0.49\linewidth]{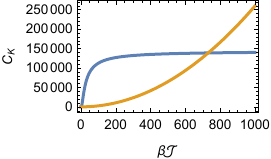}}
    \caption{Krylov operator complexity for chatic operator, $\mathcal{C}_{\rm K}(t)$, for the approximate chaotic saddle point \eqref{eq:Krylov chaotic} ({blue}) and quasi-integrable \eqref{eq:integrable Krylov} ({orange}) approximate saddle point, as a function of $\beta \mJ$ with $\kappa=0.24$ and $t=1$ (left) or $t=2$ (right). }
    \label{fig:trnasition_krylov}
\end{figure}

\subsection{Perturbative approach to Krylov operator complexity}\label{ssec:perturbative}
In this section, we investigate the limiting behavior of Krylov operator complexity when the system is very far from the phase transition line, corresponding to $\kappa\ll1$ for the chaotic and $\nu\ll1$ the quasi-integrable phases, as seen below. The analytical expressions of the complexity presented here support our numerical complexity dynamics far from the transition point as shown in Fig.\,\ref{fig:bnkcbetakappa}.

\paragraph{Chaotic saddle}
To compute the Krylov complexity, we apply Euclidean continuation in \eqref{eq:euclidean chaotic saddle} by $\tau\rightarrow \rmi t+\frac{\beta}{2}$ \cite{Blommaert:2024ydx}, which then gives:
\begin{equation}\label{eq:correlator chaotic}
    \begin{aligned}
       l=&{{-}}2\log\left[\frac{\cos\left(\frac{\pi v}{2}\right)}{\cosh\left[\frac{\pi v t}{\beta}\right]}\right]+\mathcal{O}(\kappa^4)\,,\quad \tilde{g}_z=2\kappa^2\log\left[\frac{\cos\left(\frac{\pi v}{2}\right)}{\cosh\left[\frac{\pi v t}{\beta}\right]}\right]+\mathcal{O}(\kappa^4)\,,
    \end{aligned}
\end{equation}
where $\beta$ is defined in \eqref{eq:def v}. Next, we evaluate the normalized autocorrelation function (\ref{eq:autocorre}) to $\mathcal{O}(\kappa^2)$:
\begin{equation}
    \varphi_0(t)=\frac{G_\Delta(t)}{G_\Delta(0)}=\sech^{2\Delta}\left[\frac{\pi v t}{\beta}\right]+\mathcal{O}(\kappa^2)\,.
\end{equation}
The Lanczos coefficients $b_n$ and Krylov operator complexity this type of correlator (\ref{eq:correlator chaotic}) can be found in \cite{Caputa:2021sib}
\begin{equation}\label{eq:Krylov chaotic}
	b_n\approx\frac{\pi v}{\beta}\sqrt{n(2\Delta+n-1)}\,,\quad\mathcal{C}_{\rm K}(t)\approx2\Delta\sinh^2\qty(\frac{\pi v}{\beta}t)\,.
\end{equation}
The result is in agreement with the numerical results in Fig. \ref{fig:bnkcbetakappa}, which also displays exponential growth of Krylov operator complexity when $\kappa\ll1$. This is expected for the operator growth hypothesis \cite{Parker:2018yvk}, since we have a chaotic system with a Lyapunov exponent $\pi v/\beta$ (i.e. it saturates the bound in \cite{Maldacena:2015waa} only when $v=1$ as defined in (\ref{eq:def v})).\footnote{However, given that the DSSYK has a bounded energy spectrum, the late time Krylov complexity should have late time linear growth \cite{Anegawa:2024yia}, which can be most straightforwardly shown from the one-particle chord Hilbert space in the pure DSSYK model \cite{Ambrosini:2025hvo,Ambrosini:2024sre,Aguilar-Gutierrez:2025pqp,Aguilar-Gutierrez:2025mxf}. It would be interesting to formulate particle Hilbert space in this model, which we leave for future directions.}

\paragraph{Integrable saddle}
Implementing the rotation $\tau\rightarrow \rmi t+\frac{\beta}{2}$ in \eqref{eq:int saddles} and evaluating up to order $\mathcal{O}(\frac{\nu^2}{\kappa^4(\beta \mathcal{J})^2})$, we get
\begin{equation}\label{eq:Krylov integrable}
    \varphi_0(t)=\frac{G_\Delta(t)}{G_\Delta(0)}=\exp\qty[-\Delta\left( \mathcal{J}\kappa t\right)^{2}]\,.
\end{equation}
This is essentially the autocorrelation function in a harmonic oscillator, whose Krylov complexity has been studied in \cite{Caputa:2021sib} 
\begin{equation}\label{eq:integrable Krylov}
    b_n=\mathcal{J}\kappa\sqrt{2\Delta~ n}\,, \quad \mathcal{C}_{\rm K}(t)=2\Delta\left( \mJ\kappa t\right)^{2}\,.
\end{equation}
The result is consistent with Fig.~\ref{fig:bnkcbetakappa} which also shows a parabolic growth when $\nu\ll1$. The result is due to the fact that the integrable system is essentially a harmonic oscillator.

We illustrate the behavior of Krylov operator complexity for the integrable (\ref{eq:integrable Krylov}) and chaotic (\ref{eq:Krylov chaotic}) 
 approximate saddle points, which we illustrate in Fig.\,\ref{fig:trnasition_krylov}.

\section{Operator size growth across the phase transition}\label{sec:OTOCs}

We now test the integrable-chaotic transition through the operator size growth at finite temperature, which is linearly related to an OTOC \cite{Roberts:2018mnp}. This can be computed by solving Liouville equations in the presence of twisted boundary conditions in the classical limit at all temperatures \cite{Qi:2018bje,Streicher:2019wek} (reviewed in App.\,\ref{sapp:OTOC twisted}). However, due to the interpolating Hamiltonian \eqref{eq:chaos-integrability interpolating Hamiltonian}, we have to rederive the twisted boundary conditions of the variables $g_n$ and $g_z$ of two kinds of chords from the effective action \eqref{eq:effective_action_nz}.

Based on the double-copy Hilbert space of the $N$ Majorana fermions and the left-and-right copies of operators \cite{Lin:2022rbf}, we introduce the size operator $\hat s$ which will be related to the chord number operator \eqref{eq:ell op def}
\begin{align}
    \hat s = \frac1{2}\sum_{j=1}^N \kc{1+i\psi^L_j\psi^R_j}.
\end{align}
We can define the size of an operator $O$ \cite{Roberts:2018mnp} as
\begin{align}
    s[O]
    \equiv\frac{\bra{\rm EPR} O_L^\dagger \hat s O_L\ket{\rm EPR}}{\bra{\rm EPR} O_L^\dagger O_L\ket{\rm EPR}}  \, ,
\end{align}
where $\ket{\rm EPR}$ is the maximally entangled state between the double-copy Hilbert space, defined by $(\psi_j^L+i\psi_j^R)\ket{\rm EPR}=0$. The expectation value on $\ket{\rm EPR}$ result in a normalized trace in the single-copy Hilbert space, such as the thermal partition function
\begin{align}\label{eq:partition_function}
    \bra{\rm EPR} e^{-\beta \hat H_{LR}/2}\ket{\rm EPR}=\Tr[e^{-\beta \hat H}]/\Tr[1],
\end{align}
where $\hat H_{LR}=\hat H_L+\zeta\hat H_R$, $\hat H_L\ket{\rm EPR}=\zeta \hat H_R\ket{\rm EPR}$ and $\zeta=(-1)^{p/2}$.

We will focus on the operators $e^{-\beta \hat H/2}$ and 
$
O_\beta(t) = e^{-\beta \hat H/4}\, O(t)\, e^{-\beta \hat H/4},
$
where $O$ is a normalized chaotic matter operator (defined in \eqref{eq:chord matter operator}) composed of $s_O$ Majorana fermions in each of its terms. Consequently, its operator size \cite{Qi:2019rpi} satisfies $s[O] = s_O$. 
We will consider the scaling $s_O\sim 1$ in the classical limit $\lambda\to0$ such that its scaling dimension $\Delta=\bar\lambda/\lambda=s_O/p\sim 1/p$.

To compute the size of the operator $O_\beta(0)$, we define the size generating function by introducing the twisted operator $e^{-\mu \hat s}$ in \eqref{eq:partition_function}, namely
\begin{align}\label{eq:generating_function}
    \bra{\rm EPR} e^{-\beta \hat H_{LR}/4}e^{-\mu \hat s} e^{-\beta \hat H_{LR}/4}\ket{\rm EPR},
\end{align}
which corresponds to the insertion of an twisted boundary condition in the thermal circle \cite{Qi:2018bje}. We insert the Majorana fermions $\psi^L_j$ and $\psi^R_j$ inside $\hat s$ at locations $\tau=\tfrac{3\beta}4$ and $\tfrac\beta4$ along the thermal circle $[0,\beta)$ respectively. 

In the double-scaling limit, the operator size in the disorder average is proportional to the chord number \cite{Lin:2022rbf}, namely,
\begin{align}
    \frac1p \avg{s[O]}_J\xrightarrow[\text{limit}]{\text{double-scaling}} \frac{1}{\lambda}\frac{\bra0 \tilde{\O}  \hat l \tilde{\O} \ket0}{\bra0 \tilde{\O} \tilde{\O} \ket0}\,,
\end{align}
where $\expval{\cdot}_J$ trace over the fermions and ensemble averaging over the couplings; the chord number operator $\hat l/\lambda$ in \eqref{eq:ell op def} defined in the chord space measures the number of chords going through the line connecting the locations $\tau=\tfrac\beta4,\,\tfrac{3\beta}{4}$ of the two Majorana fermions $\psi^L_j,\,\psi^R_j$ in the chord diagram. $\tilde\O$ is a operator acting on the chord space that corresponds to the operator $O$ in the SYK model.

The insertion in \eqref{eq:generating_function} results in an additional term in the effective action \eqref{eq:effective_action_nz}
\begin{align}\label{eq:new action}
    S'=S+\tilde\mu l(\tfrac{3\beta}4,\tfrac\beta4)\,,
\end{align}
where $\tilde\mu=\mu p$. So the eigenvalue of $\hat l$ is $l(\tfrac{3\beta}4,\tfrac\beta4)$ defined in \eqref{eq:l}. Following the derivation in Sec.\,\ref{ssec:correlation}, the additional term leads to the deformed Liouville equation
\begin{align}\label{eq:Liouville_twisted}
\begin{split}
    \partial_1\partial_2g_n(\tau_1,\tau_2)+2\nu^2\J^2 e^{g_n(\tau_1,\tau_2)+g_z(\tau_1,\tau_2)-\tilde\mu\theta(\tau_1,\tau_2)}&=0\,,\\
    \partial_1\partial_2g_z(\tau_1,\tau_2)+2\kappa^2\J^2 e^{g_n(\tau_1,\tau_2)-\tilde\mu\theta(\tau_1,\tau_2)}&=0\,,
\end{split}
\end{align}
where
\begin{equation}
    \theta(\tau_1,\tau_2)=\begin{cases}
        1,& \kc{\tau_1\in (-\tfrac\beta4,\tfrac\beta4) \land \tau_2\in (\tfrac\beta4,\tfrac{3\beta}4)} \lor \kc{\tau_1\in (\tfrac\beta4,\tfrac{3\beta}4) \land \tau_2\in (-\tfrac\beta4,\tfrac\beta4)},\\
        0,& \text{others},
    \end{cases}
\end{equation}
up to the periodicity $\tau\sim\tau+\beta$. 
The boundary conditions, with the corresponding symmetries from \eqref{eq:new action}, are
\begin{subequations}\label{eq:bdry cond symmetries}
    \begin{align}
    g_{n/z}(\tau,\tau)&=0,\\
    g_{n/z}(\tau_1,\tau_2)=g_{n/z}(\tau_1+\beta,\tau_2)&=g_{n/z}(\tau_1,\tau_2+\beta),\\
    g_{n/z}(\tau_1,\tau_2)=g_{n/z}^*(-\tau_2,-\tau_1)=g_{n/z}^*(\tfrac\beta2-\tau_2,\tfrac\beta2-\tau_1)&=g_{n/z}(\tau_2+\tfrac\beta2,\tau_1+\tfrac\beta2). \label{eq:reflection_symmetry}
\end{align}
\end{subequations}
The fundamental domain can then be chosen as the square 
\begin{eqnarray}\label{eq:domain}
    (0<\tau_1+\tau_2<\tfrac\beta2) \cap (0<\tau_1-\tau_2<\tfrac\beta2)\,.
\end{eqnarray} 
Once we solve \eqref{eq:Liouville_twisted} with the boundary conditions, we obtain the two-point function of $O$ in the presence of the insertion of $e^{-\mu \hat s}$. For example,
\begin{align}\label{eq:new inner prod}
    e^{-\Delta l_\mu(\beta/2,0)-\mu s_O}=\frac{\bra{\rm EPR} O_L e^{-\beta\hat H_{LR}/4}e^{-\mu \hat s} e^{-\beta\hat H_{LR}/4}O_L\ket{\rm EPR}}{\bra{\rm EPR} e^{-\beta \hat H_{LR}/4}e^{-\mu \hat s} e^{-\beta\hat H_{LR}/4}\ket{\rm EPR}}\,,
\end{align}
where the subscript ``$_\mu$'' denotes the solution of $l(\tau_1,\tau_2)$ (where $-l=g_n+g_z$) in the Liouville equation \eqref{eq:Liouville_twisted} in the presence of insertion $\rme^{-\mu \hat s}$.

To compute the OTOC or operator size along real time, we should further introduce an analytical continuation in the numerator of \eqref{eq:new inner prod}
\begin{align}
    \bra{\rm EPR} e^{-(\beta/4+it)\hat H_L-(\beta/4-it)\zeta\hat H_R}e^{-\mu \hat s} e^{-(\beta/4-it)\hat H_L-(\beta/4+it)\zeta\hat H_R}\ket{\rm EPR}\,,
\end{align}
which requires a deformed time contour $\tau(s)$ with periodicity $s\sim s+2$ and condition
\begin{align}\label{eq:contour_constraint}
\tau(0)=0,\quad 
\tau(1/2)=\tfrac\beta4+it,\quad \tau(1)=\tfrac\beta2,\quad \tau(3/2)=\tfrac{3\beta}4+it,\quad\tau(2)=\beta
\end{align}
in the effective action \eqref{eq:effective_action_nz}. The deformed Liouville equation \eqref{eq:Liouville_twisted} is covariant under the choice of time contour $\tau(s)$. In detail, 
\begin{align}\label{eq:Liouville_twisted_covariant}
\begin{split}
    \partial_{s_1}\partial_{s_2}\tilde g_n(s_1,s_2)+2\nu^2\J^2 \tau'(s_1)\tau'(s_2) e^{\tilde g_n(s_1,s_2)+\tilde g_z(s_1,s_2)-\tilde\mu\tilde\theta(s_1,s_2)}&=0\,,\\
    \partial_{s_1}\partial_{s_2}\tilde g_z(s_1,s_2)+2\kappa^2\J^2 \tau'(s_1)\tau'(s_2)e^{\tilde g_n(s_1,s_2)-\tilde\mu\tilde\theta(s_1,s_2)}&=0\,,
\end{split}
\end{align}
where $\tilde g_{n/z}(s_1,s_2)=g_{n/z}(\tau(s_1),\tau(s_2))$ and
\begin{equation}
    \tilde\theta(s_1,s_2)=\begin{cases}
        1,& \kc{s_1\in (-\tfrac12,\tfrac12) \land s_2\in (\tfrac12,\tfrac32)} \lor \kc{s_1\in (\tfrac12,\tfrac32) \land s_2\in (-\tfrac12,\tfrac12)},\\
        0,& \text{otherwise}.
    \end{cases}
\end{equation}
The symmetry \eqref{eq:bdry cond symmetries} still holds for $\tilde g_{n/z}$ after the reparametrisation. The fundamental domain \eqref{eq:domain} becomes 
\begin{eqnarray}\label{eq:domain_covariant}
    (0<s_1+s_2<1) \cap (0<s_1-s_2<1)\,.
\end{eqnarray}
By solving the Liouville equations \eqref{eq:Liouville_twisted_covariant} and getting $l_\mu(\tau(s_1),\tau(s_2))=-\tilde g_n(s_1,s_2)-\tilde g_z(s_1,s_2)$, we obtain the real-time two-point function in the presence of insertion. For example, 
\begin{align}\label{eq:OTOC}
    e^{-\Delta l_\mu(\beta/2,0)-\mu s_O}
    =\frac{\bra{\rm EPR} O_L e^{-(\beta/4+it)\hat H_L-(\beta/4-it)\zeta\hat H_R}e^{-\mu \hat s} e^{-(\beta/4-it)\hat H_L-(\beta/4+it)\zeta\hat H_R}O_L\ket{\rm EPR}}
    {\bra{\rm EPR} e^{-\beta\hat H_{LR}/4}e^{-\mu \hat s} e^{-\beta\hat H_{LR}/4}\ket{\rm EPR}}\,.
\end{align}
Because of the condition \eqref{eq:contour_constraint}, the coefficients $\tau'(s_1)\tau'(s_2)$ in the Liouville equations \eqref{eq:Liouville_twisted_covariant} are complex. In general, the solutions $\tilde g_{n/z}(s_1,s_2)$ take complex values as well. But $l_\mu(\beta/2,0)$ is always real because of the symmetry \eqref{eq:reflection_symmetry}. The change in operator size is given by
\begin{align}
    \Delta_\beta s(t) := s[O_\beta(t)]-s[e^{-\beta \hat H/2}] =s_O(1+\partial_{\tilde\mu} l_\mu(\beta/2,0)|_{\mu=0})\,.
\end{align}
To extract the scrambling characteristic in the operator size, it is sufficient to consider the relative size growth only, 
\begin{align}\label{eq:relative_size_growth}
    \delta s(t):=\frac{\Delta_\beta s(t)-\Delta_\beta s(0)}{s_O}=\partial_{\tilde\mu}l_\mu(\beta/2,0)|_{\tilde\mu=0}\,.
\end{align}

\begin{figure}
    \centering \includegraphics[height=0.285\linewidth]{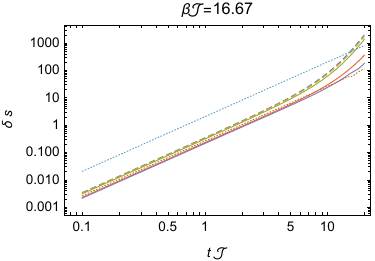} \includegraphics[height=0.285\linewidth]{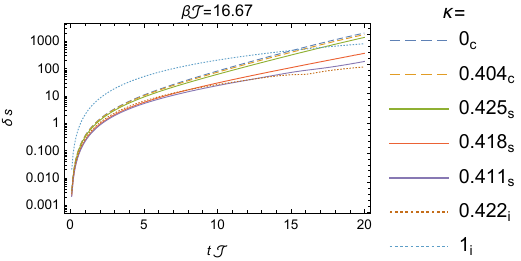}
    \caption{The relative size growth $\delta s$ as functions of dimensionless time $t\J$ at $\beta\J=16.67$ and $\kappa=0_c,\,0.404_c,\,0.425_s,\,0.418_s,\,0.411_s,\,0.422_i,\,0_i$ in the log-log plot and the log plot, where the subscripts $_c,\,_s$, and $_i$ as well as the dashed, solid, and dotted curves indicate the chaotic saddles, subdominant saddles, and quasi-integrable saddles, respectively. The first-order phase transition happens at $\kappa=0.417$.}
    \label{fig:size}
\end{figure}

We now solve the Liouville equations \eqref{eq:Liouville_twisted_covariant} numerically. To avoid discontinuous functions, we choose a smooth time contour
\begin{align}
    \tau(s)=\frac{\beta s}{2}+it\sin^2(\pi s).
\end{align}
We solve the equation using the spectral method in the square domain \eqref{eq:domain_covariant} on the variables $s_\pm=s_1\pm s_2$ and extract the derivative \eqref{eq:relative_size_growth} at $(s_+,s_-)=(1,1)$ with small values of $\tilde\mu$.

Throughout, we fix $\beta \J=16.67$ and scan $\kappa \in[0,1]$ along the three branches of saddles as shown in Fig.\,\ref{fig:action}, where the phase transition occurs at $\kappa = 0.417$.
The resulting relative size growth $\delta s(t)$ is shown in Fig.\,\ref{fig:size}. 
We observe a crossover between hyperbolic-cosine growth, 
$\delta s \propto \cosh(\lambda_L t)-1$ with $\lambda_L\approx 0.3$, 
in the branch of chaotic saddles and power-law growth, 
$\delta s \propto t^2$, 
in the branch of quasi-integrable saddles, with the branch of subdominant saddles smoothly interpolating between the two scaling.

However, when we consider the first-order phase transition between the chaotic saddle and the quasi-integrable saddle at the transition point, the growth  exhibits a discontinuous jump between hyperbolic-cosine and power-law scaling, signaling a discontinuous transition in quantum chaos. 
This is consistent with our observations obtain using the  Krylov operator complexity.

\section{Summary and outlook}\label{sec:discussion}

\subsection{Summary}

We considered the chord Hamiltonian model of \cite{Berkooz:2024evs,Berkooz:2024ofm}, which exhibits a first-order phase transition between the chaotic phase and the quasi-integrable phase at low temperature. The branch of subdominant saddles continuously interpolate between the branch of chaotic saddles and the branch of quasi-integrable saddles giving the phases. In this work, we investigate the real-time dynamics in these phases and saddles, and test their expected scrambling behavior.

First, we studied the total chord number and identified it with the Krylov state complexity in the classical limit. It grows linearly in the chaotic phase and quadratically in the quasi-integrable phase.

Second, we analyze the Lanczos coefficients and the Krylov complexity of a light chaotic matter operator. Across the chaotic, subdominant, and quasi-integrable saddles, the Lanczos coefficients $\{b_n\}$ interpolate from linear growth to square-root growth. Correspondingly, the Krylov complexity transitions from exponential growth to quadratic growth, where the latter in the quasi-integrable saddle is due to the smallness of the $b_2$ coefficient. As the branch of subdominant saddles are bypassed at the first-order transition point, the Krylov operator complexity exhibits a discontinuous jump from exponential to quadratic growth.

Third, we investigate the operator size of the chaotic matter operators. In the branches of chaotic, subdominant, and quasi-integrable saddles, the operator size exhibits exponential, intermediate, and quadratic growth, respectively. Similarly, the operator size exhibits a discontinuous jump from exponential to quadratic growth at the first-order transition.

In summary, we demonstrate that the chaotic and quasi-integrable phases exhibit distinct scrambling behaviors, which exhibit discontinuous changes at the first-order transition point. In this way, we show that also the chaos measures reflect the presence of a first-order phase transition. Below, we comment on several future directions.

\subsection{Outlook}
\paragraph{Towards the bulk dual}
As emphasized in the introduction, there are significant pieces of evidence that point to sine dilaton gravity being dual to the DSSYK model.
It was also realized in \cite{Almheiri:2024xtw} that the integrable chord theory can be interpreted as the DSSYK model at high temperatures, and it can be described by sine dilaton gravity in its flat space limit \cite{Blommaert:2024whf} (namely, flat space JT gravity \cite{Afshar:2019axx,Afshar:2021qvi}). Based on these relations, a natural expectation for the bulk dual theory of the BBJM model is that it corresponds to
\begin{equation}\label{eq:full theory}
\begin{aligned}
    I&=\nu I_{\rm SD}+\kappa {I_{\rm flat-JT}}\,,\\
I_{\rm SD}&=-\frac{1}{16\pi G_N}\qty(\int_{\mathcal{M}}\rmd^2x\sqrt{g}\qty(\Phi\mathcal{R}+2\sin\Phi)+2\int_{\partial\mathcal{M}}\rmd x\sqrt{h}\qty(\Phi_{B} K-{\rm counterterm}))\,,\\
I_{\rm flat-JT}&=-\frac{1}{16\pi G_N}\qty(\int_{\mathcal{M}}\rmd^2x\sqrt{g}\Phi\mathcal{R}+2\int_{\partial\mathcal{M}}\rmd x\sqrt{h}\qty(\Phi_{B} K-{\rm counterterm}))\,,
\end{aligned}
\end{equation}
where $\mathcal{M}$ is the manifold, $G_N$ Newton's constant, $\mathcal{R}$ the Ricci scalar, $K$ the mean curvature at $\partial\mathcal{M}$, $g_{\mu\nu}$ the metric in $\mathcal{M}$, $h_{mn}$ the induced metric in $\partial\mathcal{M}$, $\Phi_B$ the value of the dilaton at the boundary.

To be able to interpret our results holographically, we need to evaluate bulk observables, specifically minimal geodesic length connecting the asymptotic boundaries in the corresponding background solutions to the bulk theory \eqref{eq:full theory}. In sine dilaton gravity, the minimal geodesic length is indeed known to be dual to Krylov operator complexity for the DSSYK model \cite{Ambrosini:2024sre}, and similarly the
chord number in the double-scaled commuting SYK model can be interpreted in terms of a geodesic length between the asymptotic boundaries of the dual geometry \cite{Almheiri:2024xtw}. Another relation between Krylov complexity and bulk geometry has been recently put forward in a series of works relating the rate of growth of Krylov complexity with the proper radial momentum of a probe particle \cite{Caputa:2024sux,Caputa:2025dep} (see also \cite{Li:2025fqz,Aguilar-Gutierrez:2025pqp,Fan:2024iop,He:2024pox,Aguilar-Gutierrez:2025kmw,Das:2024tnw,Das:2024zuu,Jian:2020qpp}) and Krylov operator complexity of the DSSYK model with matter chords \cite{Aguilar-Gutierrez:2025kmw}. It would be interesting to develop the holographic interpretation of the corresponding observables in this work. On the other hand,  we also expect that OTOCs in the bulk theory \eqref{eq:full theory} with a scalar wordline particle would match our results in Sec.\,\ref{sec:OTOCs}. To make progress in this direction, we might apply recent lessons from \cite{Cui:2025sgy}. The authors find that OTOCs can be computed in sine dilaton gravity by evaluating end-of-the-world branes. It might be possible to implement the gluing and splitting procedure in \cite{Cui:2025sgy} for the theory \eqref{eq:full theory} with matter insertion to investigate the integrable-chaotic transition from the bulk perspective.

\paragraph{Finite N realization}
In this work, we focused on appropriate measures of chaos for strictly $N\rightarrow\infty$ systems. However, there are several useful  measures of chaos for $N\gg1$ but finite. For instance, the level spacing statistics can be used as a definition integrability, quantum chaos or something in between. The spectral form factor \cite{Cotler:2016fpe} is another measure with great relevance, which could be used to confirm some of the outcomes in this study by discerning between integrable and chaotic behaviour near the phase transition, depending how one approaches it in the phase diagram. Moreover, there are closely connected systems, such as the Rosenzweig-Porter model that also contains a chaotic Hamiltonian, i.e.~obeying random matrix statistics, and an integrable Hamiltonian that are decoupled. The connection with the DSSYK model is more explicit when we consider its $q\rightarrow0$ limit, given that it recover a similar random matrix theory behaviour \cite{Berkooz:2024lgq}. To make a meaningful comparison, one might try to pick appropriate large values for $N$ and $p$ in a single realization of the SYK that mimic those of the DSSYK model; and similarly for the Rosenzweig-Porter and BBJM models. Useful numerical results in this direction can be found e.g. \cite{Chakraborty:2025bhk,Jia:2019orl}.\footnote{We thank Micha Berkooz for useful discussions about this.} This could allow us to study the chaos-integrability transition with other chaos measures. For instance, there are other important characteristics about the Rosenzweig-Porter model that can be deduced from the inverse participation ratio in Krylov complexity, which measures the degree of localization \cite{Bhattacharjee:2024yxj}, and (multi)fractality in the system \cite{Buijsman:2021xbi}. We are preparing a detailed study about these models to highlight their connection.

 \paragraph{Local quenches} One might also consider how the two-point functions evaluated in Sec.\,\ref{sec:Krylov operators} are modified due to local quenches explicitly altering the Hamiltonian evolution at a given initial time $t=t_0$:
    \begin{eqnarray}\label{eq:transition phases}
        \ket{\psi(t)}=\begin{cases}
            \rme^{-\rmi t\eval{\hat{H}}_{\kappa=\kappa_1}}\ket{\psi_0}\,,&t\leq t_0\,,\\
            \rme^{-\rmi t\eval{\hat{H}}_{\kappa=\kappa_2}}\ket{\psi_0}\,,&t\geq t_0~  ,
        \end{cases}
    \end{eqnarray}
    where $\kappa_{1,2}$ indicates different values for $\kappa$ in \eqref{eq:transfer_mat}. One could study first and second-order phase transitions in the BBJM model under the local quench above using an appropriate order parameter, such as $\kappa$ in \eqref{eq:transfer_mat}. This could allow us to study evolution of real time thermal correlation (\ref{eq:autocorre}) and the free energy corresponding to the on-shell action \eqref{eq:effective_action_nz} in the quenched systems. There are different examples in condensed matter (see e.g. \cite{Wu_2025,Barrows:2025zos,Gomez-Ruiz:2016ggh,Swislocki:2018fje}) and holographic (see e.g. \cite{Kim:2019lxb,Rangamani:2015agy,Shimaji:2018czt}) systems where local quenches may alter macroscopic properties of the system that are similar to those in hysteresis where the properties of the system depend on its past history. It would be interesting to investigate if \eqref{eq:transition phases} indeed results in these effects in the corresponding phase transition.
    
    \paragraph{Multiple flavored chord Hamiltonians}It might be interesting to extend our study about dynamical probes of chaos for more general types of chord Hamiltonian systems (relevant discussions have been raised in \cite{Berkooz:2024ifu,Gao:2024lem}), where instead of two systems one includes arbitrarily many (see App.\,\ref{sapp:multifield}). In particular, the double-scaling limit of the commuting SYK model \cite{Gao:2023gta} with multiple types (which we refer to as ``flavors'') of chords was developed in \cite{Gao:2024lem}. In this case, the condition $q_{ij}=q\forall i,j\in\mathbb{N}$ is automatically imposed in the setup, which allows to study the collective chaotic behaviour of multiple double-scaled ``integrable'' SYKs and its thermodynamics. This would be a very convenient setting to further investigate the construction of the Krylov basis for spread complexity in Sec.\,\ref{sec:krylov basis spread} where the condition $q_{ij}=q$ leads to analytically solvable expressions. This may allow us to further develop our study of Krylov operator complexity based on the two-point functions (Sec.\,\ref{sec:Krylov operators}) using the results in \cite{Gao:2024lem}. This could extend our study of the integrable-chaotic phase transition in a chord diagram Hamiltonian with multiple (integrable) fields. We hope to report on new findings regarding this direction in the future.

\section*{Acknowledgments}
We thank Marco Ambrosini, Micha Berkooz, Andreas Blommaert and Saskia Demulder for useful discussions. We are indebted to the organizers of the Joint Belgian hep-th seminars at Université Libre de Bruxelles; the YITP-I-25-01 ``Black Hole, Quantum Chaos and Quantum Information'' workshop at Yukawa Institute for Theoretical Physics, Kyoto University; ``QIQG 2025: Quantum Information In Quantum Gravity'' at Perimeter Institute and “2025 East Asia Joint Workshop on Fields and Strings” at Okinawa Institute of
Science and Technology where this work was developed at different stages; and ``Gravity meets quantum information" at Würzburg University, which allowed this collaboration to start. SEAG thanks the High Energy Physics group in UC Santa Barbara and Rikkyo University for hosting him during different stages of this manuscript, and for travel support from the QISS consortium, and the YITP. SEAG is supported by the Okinawa Institute of Science and Technology Graduate University. This work was made possible through the support of the WOST, WithOut SpaceTime project (\hyperlink{https://withoutspacetime.org}{https://withoutspacetime.org}), supported by Grant ID\# 63683 from the John Templeton Foundation (JTF), and ID\#62312 grant from the JTF, as part of the ‘The Quantum Information Structure of Spacetime’ Project (QISS), as well as Grant ID\# 62423 from the JTF. The opinions expressed in this work are those of the author(s) and do not necessarily reflect the views of the John Templeton Foundation. RND's work leading to this publication was supported by the PRIME programme of the German Academic Exchange Service (DAAD) with funds from the German Federal Ministry of Research, Technology and Space (BMFTR).  RND and JE are also supported by Germany's Excellence Strategy through the W\"urzburg‐Dresden Cluster of Excellence on Complexity, Topology and Dynamics in Quantum Matter ‐ ctd.qmat (EXC 2147, project‐id 390858490). ZYX also acknowledges support from the berlin Quantum Initiative. 

\appendix

\section{Supplementary Background Material}\label{app:background}
In this appendix we provide complementary background to Sec.\,\ref{eq:background},. We explain about the definitions of the chaos measures used in the main text, including Krylov state (spread) complexity (App.\,\ref{sapp:spread}), Krylov operator complexity (App.\,\ref{sapp:KRylov op}), OTOCs at finite temperatures (App.\,\ref{sapp:OTOC twisted}); basics on the DSSYK model, as well as the commuting DSSYK model (App.\,\ref{sapp:DSSYK basic}), which is a specific realization of the integrable model $\hH_2$ in \eqref{eq:chaos-integrability interpolating Hamiltonian}; and the multifield formalism for multiple ensemble-averaged double-scaled model (App.\,\ref{sapp:multifield}).

\subsection{Krylov state (spread) complexity}\label{sapp:spread}
Starting from the Schrödinger picture for a generic pure quantum system, we would like to construct an ordered, orthonormal basis of states $\qty{\ket{B_n}}$ that minimizes $$\sum_n c_n\abs{\bra{\phi(t)}\ket{B_n}}^2$$ where c$_n$ is an arbitrary monotonically increasing real sequence, and
\begin{equation}
     \ket{\phi(t)}=\rme^{-\rmi Ht}\ket{\phi_0}\,.
\end{equation}
It was found in \cite{Balasubramanian:2022tpr} that the solution to this problem is the so-called Krylov basis, $\ket{K_n}$, defined through the Lanczos algorithm shown below
\begin{align}
    \ket{A_{n+1}}&\equiv (\hat{H} - a_n)\ket{K_n} - b_n \ket{K_{n-1}}\,, \label{eq:lanczos}\\
    \ket{K_n} &\equiv b_n^{-1}\ket{A_n}\,.
\end{align}
Here $\ket{K_0}\equiv\ket{\phi_0}$ and
\begin{equation}
    a_{n} \equiv \bra{K_n} \hat{H} \ket{K_n}, \qquad b_{n} \equiv(\braket{A_n}{A_n})^{1/2}\,,
\end{equation}
are called the Lanczos coefficients. Using this basis, $\ket{\phi(t)}$ can be expressed as
\begin{equation}\label{eq: phi (t) spread}
    \ket{\phi(t)} = \sum^{\mathcal{K}}_{n=0} \phi_n (t) \ket{K_n}\,.
\end{equation}
Here $\mathcal{K}$ denotes the Krylov space dimension, which satisfies $\mathcal{K}\leq {\rm dim}(\mathcal{H})$. The Hamiltonian in this basis becomes tridiagonal, and we can express a recursive relation between the time-dependent components in (\ref{eq: phi (t) spread}) as a Schrödinger equation:
\begin{equation} \label{eq:Schro}
    \rmi\partial_t \phi_n(t) = a_n \phi_n (t) + b_{n+1}\phi_{n+1}(t) + b_n \phi_{n-1}(t)\,,
\end{equation}
with $\sum_n|\phi_n(t)|^2=1$. Krylov complexity for states (also called spread complexity) can be defined as
\begin{equation} \label{eq:S Complexity}
   \mathcal{C}_{\rm S}(t) \equiv \sum_n f(n) |\phi_n(t)|^2 \,,
\end{equation}
where $f(n)$ is a monotonically increasing sequence in terms of $n$. In the main text, we set $f(n)=n$, so that spread complexity represents a position expectation value in an ordered lattice given by the Krylov basis $\ket{K_n}$. 

Intuitively, $\mathcal{C}_{\rm S}$ measures the average position in a one-dimensional chain generated by the Krylov basis, where each step along the chain represents an increasingly chaotic state since they roughly behave as $\ket{K_n} \approx \hat{H}^n \ket{\phi_0}$.

\subsection{Krylov operator complexity}\label{sapp:KRylov op}
Consider an operator $\hat{O}$ in the Heisenberg picture. We would like to express it in terms of a complete and ordered basis of states $\qty{\ket{\chi_n}}$ as
\begin{equation}\label{eq:Op as state Krylov}
	\begin{aligned}
		|O)&\equiv\sum_{m,\,n}O_{nm}\ket{\chi_m,~\chi_n}\,,
	\end{aligned}
\end{equation}
where $O_{nm}\equiv\bra{\chi_m}\hat{O}\ket{\chi_n}$. Given the thermal ensemble describing the DSSYK model, we will define the Frobenious inner product (while for thermal ensembles one needs to modify the definition, see e.g. \cite{Parker:2018yvk,Barbon:2019wsy,Anegawa:2024yia}): 
\begin{equation}\label{eq:inner prod}
	(X|Y)=\frac{1}{Z(\beta)}\tr(\rme^{-\frac{\beta}{2}\hat{H}}\hat{X}^\dagger\rme^{-\frac{\beta}{2}\hat{H}} \hat{Y})\,.
\end{equation}
We can represent the evolution of the operator through the Heisenberg equation as
\begin{align}\label{eq: Heisenberg time evol}
	\partial_t|O(t))&=\rmi\hL|O(t))\,,
\end{align}
where $\mathcal{L}$ is called the Liouvillian super-operator,
\begin{equation}
	\hat{\mathcal{L}}=\qty[\hat{H},~\cdot~],\quad \hat{O}(t)=\rme^{\rmi\hL t}\hat{O}\,.
\end{equation}
We can then solve (\ref{eq: Heisenberg time evol}) in terms of a Krylov basis, $\qty{|O_n)}$,
\begin{equation}\label{eq:amplitudes}
	\begin{aligned}
		|O (t))&=\sum_{n=0}^{\mathcal{K}-1} i^n \varphi_n(t) |O_n)\,,\\
		\varphi_n(t)&=(O_n|\rme^{\rmi \hat{\mathcal{L}}t}|O_n)\,,\quad (O_m|O_n)=\delta_{mn}\,.
	\end{aligned}
\end{equation}
Moreover, assuming that $\hat{O}(t)$ is a Hermitian operator, the correlation function is an even function in $t$ that can be expanded as a Taylor series as
\begin{equation}\label{eq:2pnt correlator Krylov}
	\varphi_0(t)=(O(t)|O(0))=\sum_nm_{2n}\frac{(-1)^{n}t^{2n}}{(2n)!}\,,
\end{equation}
where $m_{2n}$ are referred to as the moments. It follows that given an autocorrelation function ($\varphi_0(t)$), the moments are easily accessed from
\begin{eqnarray}\label{eq:moment_alt}
    m_{2n}=\eval{\qty(-1)^n\qty(\dv{t})^{2n}\varphi_0(t)}_{t=0}\,.
\end{eqnarray}
The Lanczos coefficients $b_n$ can be then determined from the moments using an algorithm \cite{Parker:2018yvk,viswanath1994recursion,Bhattacharjee:2022ave}
\begin{align}\label{eq:Alt Lanczos}
	b_n=\sqrt{Q_{2n}^{(n)}}\,,\quad Q_{2k}^{(m)}=\frac{Q_{2k}^{(m-1)}}{b_{m-1}^2}-\frac{Q_{2k-2}^{(m-2)}}{b_{m-2}^2}\,,
\end{align}
where $Q_{2k}^{(0)}=m_{2k}$, and $Q_{2k}^{(-1)}=0$.

The other amplitudes can be determined through the Lanczos algorithm and the Heisenberg equation (\ref{eq: Heisenberg time evol}), leading to the recursion relation:
\begin{equation}\label{eq:sch eq K operator}
	\partial_t\varphi_n(t)=b_n\varphi_{n-1}(t)-b_{n+1}\varphi_{n+1}(t)\,.
\end{equation}
Krylov operator complexity is then defined as
\begin{equation}\label{eq:Krylov complexity}
	{\mathcal{C}_{\rm K}}(t)\equiv\sum_{n=0}^{\mathcal{K}-1}n|\varphi_n(t)|^2\,.
\end{equation}
The definition above was originally motivated \cite{Parker:2018yvk} to describe the size of the operator under Hamiltonian evolution, as it measures the mean width of a wavepacket in the Krylov space.

\subsection{OTOCs at finite temperature}\label{sapp:OTOC twisted}
In connection to Sec.\,\ref{sec:OTOCs}, we now review how to carry out the evaluation of crossed four point functions at finite temperature in the SYK model, based on the \cite{Qi:2018bje,Streicher:2019wek}.

In the DSSYK model, a semiclassical approximation for OTOCs away from the triple-scaling limit was derived in \cite{Aguilar-Gutierrez:2025pqp,Aguilar-Gutierrez:2025mxf} (see also e.g. \cite{Lin:2023trc,Narovlansky:2025tpb}), while the triple scaling limit (i.e. the low-energy regime) result had been obtained in previous works \cite{Berkooz:2022fso,Berkooz:2018jqr}. 

However, once we incorporate integrable and chaotic chords, the procedure above requires further adjustments. We choose to approach this problem with from the generating function of general fermionic correlation functions at finite temperature (first illustrated in \cite{Qi:2018bje,Streicher:2019wek}), and then take the double-scaling limit of the results. For this reason we begin with the $2n$-point generating function
\begin{eqnarray}\label{eq:Zcorrelator}
    Z_\mu[\rho^{1/2}]=\bra{\rho^{1/2}}\rm e^{-\mu \hat s}\ket{\rho^{1/2}}\,,
\end{eqnarray}
where the factor $\rme^{-\mu\hat s}$ generates Euclidean evolution. The crossed four-point function of interest is
\begin{equation}
    \mathcal{G}_\mu(\tau_a,\tau_b)=\frac{\bra{0}\mathcal{T}\rme^{-\frac{\beta}{2}\qty(\hH_L+\hH_R)}\rme^{-\mu \hat{s}\qty(\frac{\beta}{4})}\psi_1^L(\tau_a)\psi_1^L(\tau_b)\ket{0}}{\bra{0}\mathcal{T}\rme^{-\frac{\beta}{2}\qty(\hH_L+\hH_R)}\rme^{-\mu n\qty(\frac{\beta}{4})}\ket{0}}\,.
\end{equation}
where $\hat{s}$ is the size operator
\begin{eqnarray}
    \hat{s}=-\frac{\rmi}{2}\sum_{j=1}^N\psi_j^L\psi_j^R\,.
\end{eqnarray}
We can see that $\mathcal{G}_{\mu=0}$ gives a two-point function, while the OTOC can be computed from $ -\lim_{\mu=0}\partial_\mu G_\mu$. This can be more conveniently evaluated using
\begin{eqnarray}\label{eq:action psiLR}
   \rme^{\mu n}\begin{pmatrix}
       \psi_L\\
       \rmi\psi_R
   \end{pmatrix}\rme^{-\mu n}=\begin{pmatrix}
        \cosh\mu&-\sinh\mu\\
        -\sinh\mu&\cosh\mu
   \end{pmatrix}\begin{pmatrix}
       \psi_L\\
       \rmi\psi_R
   \end{pmatrix}\,,
\end{eqnarray}
which can be expressed in terms of a single field $\psi$ defined as
\begin{eqnarray}
    \psi_i(\tau)=\begin{cases}
        \psi_L(\tau)\,,&0<\tau<\beta/2\,,\\
        \psi_R(\beta-\tau)\,,&\beta/2<\tau<\beta\,,\\
    \end{cases}
\end{eqnarray}
where we impose twisted boundary conditions, defined by
\begin{eqnarray}\label{eq:twisted bdry}
    \psi(\tau)=-\psi(\tau+\beta)\,,
\end{eqnarray}
so that \eqref{eq:action psiLR} can be expressed
\begin{eqnarray}\label{eq:matrix mu}
     \lim_{\mu\rightarrow\beta/4^+}\rme^{\mu n}\begin{pmatrix}
       \psi(\tau)\\
       \psi(\beta-\tau)
   \end{pmatrix}\rme^{-\mu n}=\lim_{\mu\rightarrow\beta/4^-}\begin{pmatrix}
        \cosh\mu&-\sinh\mu\\
        -\sinh\mu&\cosh\mu
   \end{pmatrix}\begin{pmatrix}
       \psi(\tau)\\
       \psi(\beta-\tau)
   \end{pmatrix}\,.
\end{eqnarray}
The role of \eqref{eq:twisted bdry} is to generate the crossing between the fermions in the OTOC.

We note that \eqref{eq:matrix mu} can be evaluated from \eqref{eq:Zcorrelator} as 
\begin{eqnarray}
   \mathcal{G}_\mu(\OD)=\frac{Z_\mu[\OD\rho^{1/2}]}{Z_\mu[\rho^{1/2}]}\,,
\end{eqnarray}
when $\OD=\psi_1(t)\rho^{1/2}$. This calculation can be expanded in terms of twisted boundary conditions:
\begin{equation}
    \begin{pmatrix}
        \lim_{\tau_{1/2}\rightarrow\beta/4^+}\mathcal{G}_\mu(\tau_1,\tau_2)\\
        \lim_{\tau_{1/2}\rightarrow\beta/4^-}\mathcal{G}_\mu(\tau_1,\tau_2)
    \end{pmatrix}=\begin{pmatrix}
        \cosh\mu&-\sinh\mu\\
        -\sinh\mu&\cosh\mu
    \end{pmatrix}\begin{pmatrix}
        \lim_{\tau_{1/2}\rightarrow\beta/4^-}\mathcal{G}_\mu(\tau_1,\tau_2)\\
        \lim_{\tau_{1/2}\rightarrow\beta/4^+}\mathcal{G}_\mu(\tau_1,\tau_2)
    \end{pmatrix}\,.
\end{equation}
$\mathcal{G}_\mu$ can then be now solved in terms of a fundamental domain where $0<\tau_1-\tau_2<\beta/2$. In our notation, when $p$ is fixed, and $N\gg1$, one can analytically recover the two-point functions and OTOCs \cite{Qi:2018bje}. We implement this in Sec.\,\ref{sec:OTOCs} for the integrable-chaotic chord Hamiltonian model.

\subsection{The Double-Scaled SYK and Commuting SYK Models}\label{sapp:DSSYK basic}
We presented the physical SYK origins of the model in Sec.\,\ref{ssec:chord Hilbert space}, we now describe its auxiliary Hilbert space in more detail. As reviewed e.g.~\cite{Berkooz:2024lgq}, one can construct an orthonormal basis (i.e. $\braket{n}{m}=\delta_{nm}$), to write the Hamiltonian of the auxiliary system in the form
\begin{equation}\label{eq:DSSYK Hamiltonian}
    \hat{H}_{1}=\frac{J}{\sqrt{\lambda}}(\hat{a}_{n}+\hat{a}_{n}^\dagger)\,,
\end{equation}
where $\hat{a}_n$ and $\hat{a}_n^{\dagger}$ are creation and annihilation operators acting on the $\qty{\ket{n}}$ basis,
\begin{equation}
    \hat{a}_n\ket{n}=\sqrt{[n]_q}\ket{n-1}\,,\quad \hat{a}^\dagger_n\ket{n}=\sqrt{[n+1]_q}\ket{n+1}\,,\quad [n]_q\equiv\frac{1-q^n}{1-q}\,.
\end{equation}
We also introduce matter chord operators in this model
\begin{eqnarray}\label{eq:chord matter operator}
    V:=\sum_{I'}J_{I'}\Psi_{I'}\,,
\end{eqnarray}
where, similarly to \eqref{eq:SYK Hamiltonian}, $I'$ is a collective indicating $1\leq i_1\leq i_2\leq\dots\leq i_{p'}\leq N$, where $\Delta:=p'/p$ is associated as the conformal dimension of the operator \cite{Berkooz:2018jqr}, and $J_{I'}:=J_{i_1,\dots,~i_{p'}}$ .

It has been noticed that the chord number basis of the DSSYK model forms a Krylov basis \cite{Lin:2022rbf}, so that $\ket{K_n}=\ket{n}$, and the spread complexity for this model has been calculated in the natural TFD state $\rme^{-\rmi t\hat{H}_1}\ket{0}$ by \cite{Rabinovici:2023yex}. The analytic expression was generalized including finite temperature effects by considering the Krylov complexity of the Hartle-Hawking state in the semiclassical regime of the DSSYK model \begin{equation}\label{eq:sonner}
   \lim_{\lambda\rightarrow0} {C_{\rm S}(t)=\frac{2}{\lambda}\log\frac{\cosh\qty(2\mJ\sin\theta~t)}{\sin\theta}\,,}
\end{equation}
where $\theta$ parametrizes the energy of the DSSYK model
\begin{eqnarray}
    E(\theta)=\frac{2\mJ}{\sqrt{\lambda(1-q)}}\cos\theta\,.
\end{eqnarray}
Further generalizations in Krylov complexity/wormhole length dictionary have been reported with matter contributions \cite{Xu:2024gfm,Ambrosini:2024sre,Aguilar-Gutierrez:2025mxf,Aguilar-Gutierrez:2025pqp}, supersymmetry \cite{Aguilar-Gutierrez:2025sqh}, conserved charges \cite{Forste:2025gng}, and finite $N$ effects \cite{Miyaji:2025ucp,Balasubramanian:2024lqk,Nandy:2024zcd}.

\paragraph{Commuting DSSYK model}
Similarly, for the commuting SYK \eqref{eq:Hamiltonian integrable} in the double scaling limit has a span of states $\qty{\ket{z}}$ are described with a Hamiltonian
\begin{equation}
    \hat{H}_2=J\qty(\ha_z+\ha^\dagger_z)\,,
\end{equation}
which now obeys a Heisenberg-Weyl algebra $[a_z,a^\dagger_z]_1=1$.

As in the previous subsection, we will consider the Krylov complexity as the expectation value of the chord number operator $\hat{s}$ for the infinite-temperature TFD state $\rme^{-\rmi t \hat{H}_{\rm com}}\ket{0}$, where $\hat{H}_{\rm com}$ is the chord Hamiltonian $\hH_2$ in (\ref{eq:Hamiltonian integrable}). The result has been computed for us by \cite{Almheiri:2024xtw} (see their (2.15), which is at finite temperature),
\begin{equation}
    \mathcal{C}_{\rm S}(t)=\frac{\mJ^2}{2}\qty(t^2+\frac{\beta^2}{4})\,.
\end{equation}
We also know the OTOC \cite{Almheiri:2024xtw} (3.8)
\begin{eqnarray}
    \begin{aligned}
&{\frac{1}{Z}} \tr\left[\rme^{-\beta H} [W(0),V(t)]^2\right] \\
&= -2  + 2  \cos \left[ {(1 - e^{-\Delta_V })(1 - e^{-\Delta_W })J^2 \beta t / 2 }\right] \,  e^{-\Delta_V \Delta_W -(1 - \rme^{-\Delta_V })(1 - \rme^{-\Delta_W }) J^2t^2}\,,
\end{aligned}
\end{eqnarray}
where $W(t)$ and $V(0)$ are matter operators \eqref{eq:chord matter operator} with conformal weight $\Delta_V$ and $\Delta_W$ respectively. 

Moreover, given that the integrable model is described by a quantum harmonic oscillator Hamiltonian, the Krylov complexity for integrable matter operators $V$ \eqref{eq:chord matter operator} with a conformal weight $\Delta$ becomes (see the Heisenberg-Weyl algebra case in \cite{Caputa:2021sib})
\begin{equation}
    \mathcal{C}_{\rm O}(t)=2\Delta (\mJ\kappa t)^2\,.
\end{equation}

\subsection{Multifield Liouville formulation}\label{sapp:multifield}
One can build other generalizations of the above double-scaled DSSYK model and its integrable counterpart, which are conveniently treated in the multifield formalism in \cite{Berkooz:2024ifu}; valid for arbitrarily many types of chords. The starting point in this formalism is the theory
\begin{eqnarray}
    \hH=\sum_{i=1}^K\kappa_{i}^2\hH_i\,,
\end{eqnarray}
where each $\hH_i$ is a random Hamiltonian in the same universality class as the original DSSYK model \cite{Berkooz:2018jqr,Berkooz:2018qkz}, and the chord crossing weight between Hamiltonians $\hH_i$ and $\hH_j$ is denoted $q^{\alpha_{ij}}$; $\kappa_i$ are numerical coefficients that we take to satisfy a convenient normalization
 \begin{eqnarray}
     \sum_{i=1}^K\abs{\kappa_i}^2=1\,,
 \end{eqnarray}
 and $K$ is the total number of random Hamiltonians. The partition function in this theory can be expressed
\begin{eqnarray}
   \int \left(\prod_{i=1}^K \mathcal{D}[g_{i}]\right) \exp\left[\frac{1}{\lambda}\iint_{0}^\beta d\tau_{1}d\tau_{2}\left[\sum_{i,j=1}^K\alpha_{ij}g_{i}\partial_{1}\partial_{2}g_{j} + 4 \mJ^2\sum_{i=1}^K\kappa_{i}^2  \rme^{\sum_{j}\alpha_{ij}g_{j}}\right] \right] \,,
\end{eqnarray}
where $\alpha_{ij}$ is a set of constants. The two-chord Hamiltonian system corresponds to
\begin{equation}
    \alpha_{ij} = \begin{pmatrix}
        1 & \Delta \\
        \Delta & \alpha' \\
    \end{pmatrix}\,.
\end{equation}
The integrable deformation of the DSSYK model \eqref{eq:chaos-integrability interpolating Hamiltonian} is then $\alpha'=0$; which is a case of interest for us.

\section{More details about \texorpdfstring{\eqref{eq:meaning r}}{}}\label{app:alt exp}
In this appendix, we study whether (\ref{eq:meaning r}) can be used as a Krylov basis for the Hamiltonian (\ref{eq:transfer_mat}). We consider the following infinite-temperature TFD state \cite{Rabinovici:2023yex,Lin:2022rbf} as a reference state
\begin{eqnarray}
    \ket{\phi(t)}:=\rme^{-\rmi \hat{H}t}\ket{0}=\sum_{n,z}\phi_{n,z}(t)\ket{n,z;~\vec{r}}\,,
\end{eqnarray}
where we have assumed $\qty{\ket{n,z;~\vec{r}}}$ is a complete basis; and we defined $\phi_{n,z}:=\llangle n,z;\vec{r}|\rme^{-\rmi\hH t}\ket{0}$. Acting with the creation/annihilation operators in (\ref{eq:transfer_mat}) on the above state results in
\begin{equation}
\begin{aligned}
    \hat{H}    \ket{\phi(t)}=&\nu\sum_{n,z}\qty[\sqrt{[n+1]_{q_n}}\phi_{n+1,z}\ket{n,z;~\vec{r}_1}+\sqrt{[n]_{q_n}}\phi_{n-1,z}\ket{n,z;~\vec{r}_2}]\\
&+\kappa\sum_{n,z}\qty[\sqrt{[z+1]_{q_z}}\phi_{n,z+1}\ket{n,z;~\vec{r}_3}+\sqrt{[z]_{q_z}}\phi_{n,z-1}\ket{n,z;~\vec{r}_4}]\,,
\end{aligned}
\end{equation}
where, given that the vectors $\vec{r}_{i}$ are manifestly different from each other (due to the different applications of $\hat{a}_i$), the basis $\qty{\ket{n,z;~\vec{r}}}$ even if complete, is not a Krylov basis in the sense of the Lanczos algorithm (see \eqref{eq:Schro}). For this reason, we focused on the special case $q_{nn}=q_{zz}=q_{nz}$ in the main text.

\section{Deviations from the \texorpdfstring{$q_{nn}=q_{nz}=q_{nz}$}{} Krylov basis}\label{ssec:deviations krylov basis}
In this appendix we complete our discussion of Sec.\,\ref{ssec:Krylov basis for all q} regarding corrections to the Krylov basis away from the case where the Hamiltonian crossings are $q_{nn}=q_{nz}=q_{zz}$.

We begin considering small derivations from $\abs{q_{nn}-q_{nz}}\sim \abs{q_{nz}-q_{zz}}\sim dz\ll 1$ and we denote the total Hamiltonian with a deviation $d$ as $H_d$. From (\ref{eq:KrylovIdenticalChord}) we identify
\begin{align}
    \delta b_n^2=\frac{n(n-1)}2 d\,. \label{eq:bn_var}
\end{align}
The variation of $\mu_{2n}$ can also be evaluated from \eqref{eq:bn_var} and \eqref{eq:deviations} as
\begin{equation}
\ln \mu_{2n} \;=\; \sum_{m=1}^{n} \ln b_m^{2},\qquad
\delta \mu_{2n}
      \;=\;
      \mu_{2n} \sum_{m=1}^{n} \frac{\delta b_m^{2}}{b_m^{2}}
      \;=\;
      \mu_{2n}\, d \sum_{m=1}^{n} \frac{m(m-1)}{2\, b_m^{2}}.
\end{equation}
Next, we define the error state
\begin{align}
    \ket{\E_{n+1}}:=H_d\ket{P_{n}} - b_n^2\ket{P_{n-1}}-b_{n+1}\ket{P_{n+1}}\,,
\end{align}
where $\ket{P_k}$ is taken as \eqref{eq:KrylovIdenticalChord}.
The relative error can then be expressed as 
\begin{align}
    \frac{\avg{\E_{n+1}|\E_{n+1}}}{\avg{P_{n+1}|P_{n+1}}}
    =\frac{\bra{P_n}(H_d-H_0)^2\ket{P_n}}{\avg{P_n|P_n}b_{n+1}^2}\,.
\end{align}
This indicates the degree to which we can approximately describe the system away from $q_{nn}=q_{nz}=q_{zz}$ with the same Krylov basis in Sec.\,\ref{ssec:special case}.

\bibliographystyle{JHEP}
\bibliography{references.bib}
\end{document}